\documentclass[fleqn,usenatbib]{mnras}

\usepackage{newtxtext,newtxmath}
\usepackage{rotating}
\usepackage{adjustbox}
\usepackage[T1]{fontenc}

\DeclareRobustCommand{\VAN}[3]{#2}
\let\VANthebibliography\thebibliography
\def\thebibliography{\DeclareRobustCommand{\VAN}[3]{##3}\VANthebibliography}

\usepackage{graphicx}	% Including figure files
\usepackage{amsmath}	% Advanced maths commands
\usepackage{xcolor}
\def \MSUN{M_{\odot}}

\definecolor{purple}{rgb}{0.5,0,0.87}

\definecolor{lightblue}{rgb}{0.1,0.5,0.89}

\definecolor{cobalt}{rgb}{0.0, 0.28, 0.67}

\definecolor{orange}{rgb}{1, 0.67, 0.0}

\definecolor{emerald}{rgb}{0.4,0.66,0.31}
 
\title[Measuring Intracluster Light]{Preparing for low surface brightness science with the Vera C. Rubin Observatory: A Comparison of Observable and Simulated Intracluster Light Fractions}

\author[S. Brough et al]{Sarah Brough,$^{1}$\thanks{E-mail: s.brough@unsw.edu.au}
Syeda Lammim Ahad,$^{2}$
Yannick M.~Bah\'e,$^{3,2,4}$
Ama\"{e}l Ellien,$^{5}$
Anthony H. Gonzalez,$^{6}$
\newauthor Yolanda Jim\'enez-Teja,$^{7,8}$
Lucas C. Kimmig,$^{9}$
Garreth Martin,$^{10,11}$
Cristina Mart\'inez-Lombilla,$^{1}$
\newauthor Mireia Montes,$^{12,13,14}$
Annalisa Pillepich,$^{15}$
Rossella Ragusa,$^{16,17}$
Rhea-Silvia Remus,$^{9}$
\newauthor Chris A. Collins,$^{18}$
Johan H. Knapen,$^{12,13}$ 
J. Christopher Mihos, $^{19}$
\\
$^{1}$School of Physics, University of New South Wales, NSW 2052, Australia\\
$^{2}$Leiden Observatory, Leiden University, P.O.~Box 9513, 2300 RA Leiden, The Netherlands\\
$^{3}$Institute of Physics, Laboratory of Astrophysics, Ecole Polytechnique F\'{e}d\'{e}rale de Lausanne (EPFL), Observatoire de Sauverny, 1290 Versoix, Switzerland\\
$^{4}$Institute for Computational Cosmology, Durham University, South Road, Durham DH1 3LE, UK\\
$^{5}$Anton Pannekoek Institute for Astronomy \& GRAPPA, University of Amsterdam, Science Park 904, 1098 XH, Amsterdam, The Netherlands\\
$^{6}$Department of Astronomy, University of Florida, Gainesville, FL 32611, USA\\
$^{7}$Instituto de Astrof\'{i}sica de Andaluc\'{i}a–CSIC, Glorieta de la Astronom\'{i}a s/n, E–18008 Granada, Spain\\
$^{8}$Observat\'orio Nacional - MCTI (ON), Rua Gal. Jos\'e Cristino 77, S\~{a}o Crist\'ov\~{a}o, 20921-400, Rio de Janeiro, Brazil\\
$^{9}$Universit\"ats-Sternwarte M\"unchen, Scheinerstr.\ 1, D-81679 M\"unchen, Germany\\
$^{10}$Korea Astronomy and Space Science Institute, 776 Daedeokdae-ro, Yuseong-gu, Daejeon 34055, Korea\\
$^{11}$Steward Observatory, University of Arizona, 933 N. Cherry Ave, Tucson, AZ 85719, USA\\
$^{12}$Instituto de Astrof\'{\i}sica de Canarias, c/ V\'{\i}a L\'actea s/n, E-38205 - La Laguna, Tenerife, Spain \\
$^{13}$Departamento de Astrof\'isica, Universidad de La Laguna, E-38205 - La Laguna, Tenerife, Spain  \\
$^{14}$Space Telescope Science Institute, 3700 San Martin Drive, Baltimore, MD 21218, USA \\
$^{15}${Max-Planck-Institut f{\"u}r Astronomie, K{\"o}nigstuhl 17, 69117 Heidelberg, Germany}\\
$^{16}$INAF-Astronomical Observatory of Capodimonte, Salita Moiariello 16, 80131 Naples, Italy \\
$^{17}$University of Naples “Federico II”, Via Cinthia 21, Naples 80126, Italy\\
$^{18}$Astrophysics Research Institute, Liverpool John Moores University, IC2, Liverpool Science Park, 146 Brownlow Hill, Liverpool L3 5RF, United Kingdom\\
$^{19}$Department of Astronomy, Case Western Reserve University, 10900 Euclid Avenue, Cleveland, OH 44106, USA}

\date{Accepted XXX. Received YYY; in original form ZZZ}

\pubyear{2023}

\begin{document}
\label{firstpage}
\pagerange{\pageref{firstpage}--\pageref{lastpage}}
\maketitle

\begin{abstract}
Intracluster Light (ICL) provides an important record of the interactions galaxy clusters have undergone. However, we are limited in our understanding by our measurement methods. To address this we measure the fraction of cluster light that is held in the Brightest Cluster Galaxy and ICL (BCG+ICL fraction) and the ICL alone (ICL fraction) using observational methods (Surface Brightness Threshold-SB, Non-Parametric Measure-NP, Composite Models-CM, Multi-Galaxy Fitting-MGF) and new approaches under development (Wavelet Decomposition-WD) applied to mock images of 61 galaxy clusters ($14<$log$_{10}M_{200c}/M_{\odot}<14.5$) from four cosmological hydrodynamical simulations. We compare the BCG+ICL and ICL fractions from observational measures with those using simulated measures (aperture and kinematic separations). The ICL fractions measured by kinematic separation are significantly larger than observed fractions. We find the measurements are related and provide equations to estimate kinematic ICL fractions from observed fractions. The different observational techniques give consistent BCG+ICL and ICL fractions but are biased to underestimating the BCG+ICL and ICL fractions when compared with aperture simulation measures. Comparing the different methods and algorithms we find that the MGF algorithm is most consistent with the simulations, and CM and SB methods show the smallest projection effects for the BCG+ICL and ICL fractions respectively. The Ahad (CM), MGF and WD algorithms are best set up to process larger samples, however, the WD algorithm in its current form is susceptible to projection effects. We recommend that new algorithms using these methods are explored to analyse the massive samples that Rubin Observatory's Legacy Survey of Space and Time will provide.
\end{abstract}

% Select between one and six entries from the list of approved keywords.
% Don't make up new ones.
\begin{keywords}
Galaxies: clusters: general -- Galaxies: haloes -- Galaxies: evolution -- Galaxies: photometry
\end{keywords}

%%%%%%%%%%%%%%%%%%%%%%%%%%%%%%%%%%%%%%%%%%%%%%%%%%

%%%%%%%%%%%%%%%%% BODY OF PAPER %%%%%%%%%%%%%%%%%%

\section{Introduction}
A diffuse collection of stars is observed to sprawl across the central regions of galaxy groups and clusters. This is the Intracluster Light (ICL), an important fossil record of all the interactions these systems have undergone. A robust understanding of the ICL, therefore, serves as a powerful probe of the evolution of cosmic structure and the build-up of the largest bound structures in the Universe. Its physical scale is similar to the scale of the dark matter in clusters, making the ICL an important potential luminous tracer of dark matter in these systems (e.g. \citealt{MT19,Deason2021,MT22,Diego2023}).

The fraction of the cluster light that is held in the ICL (ICL fraction), and its dependence on cluster mass and redshift, are important tools in understanding how galaxies and clusters evolve. However, there exists a crucial problem with the use of ICL for this science: the ambiguous observational definition of the ICL. The ICL is observed to be concentrated around the cluster's massive central galaxy (Brightest Cluster Galaxy or BCG).  Deep images of clusters of galaxies show that the transition between the BCG and the ICL happens smoothly with no clear break point. So, without any information about the kinematics of the stars, separating the ICL contribution from that of the BCG is challenging.  Clusters also contain satellite galaxies which contribute their own diffuse light to the ICL.  As a result, observers have developed a range of techniques to measure the ICL fraction. These include the following:

(i) The Surface Brightness (SB) Threshold method \citep[e.g.,][]{Feldmeier2004,Presotto2014, MT14,Burke2015, MT18, Montes2021, Furnell2021, Martinez-Lombilla2023a} considers all light below a certain surface brightness threshold to be part of the ICL. While this method is simple to apply and does include light around satellite galaxies, it does not capture the ICL projected over the BCG and the satellite galaxies.  In addition, observations with different depths lead to different ICL fractions and observers use different photometric bands (due to the limited availability of deep images, or different redshifts), and different thresholds. Although this method is easy to apply, these caveats make it very difficult to compare results between studies. 

(ii) Non-Parametric Measure (NP) \citep[e.g.,][]{Gonzalez2007,DeMaio2018,Martinez-Lombilla2023a} this method measures the BCG+ICL fraction without making any assumption regarding the shape of the BCG or ICL distribution. This method does capture the ICL projected over the BCG and can potentially capture the diffuse light associated with satellite galaxies.

(iii) The Composite Model (CM) method combines different empirical models, normally a double S\'ersic model \citep{Sersic1968} or a S\'ersic and an exponential \citep[e.g.,][]{Gonzalez2005, Seigar2007,Presotto2014, Iodice2016,Spavone2017b, MT18, RAGUSA2021,ragusa2023, Montes2021, Martinez-Lombilla2023a, Joo2023,Ahad2023} to define and separate the BCG and the ICL. This method does capture the ICL projected over the BCG but the choice of model parameters and the intrinsic difficulty of the problem means that this method can be very degenerate \citep{janowiecki2010}. It also fails to capture the diffuse light associated with satellite galaxies.

(iv) Multi-Galaxy Fitting (MGF) methods model and remove all the galaxies in the image with either traditional analytical profiles \citep[e.g.][]{Giallongo2014, Morishita2017, Poliakov2021} or orthonormal mathematical bases \citep{Jimenez-Teja2016, Jimenez-Teja2018}. These methods, along with Wavelet Decomposition, have the advantage of separating galaxies and ICL for the whole image, thereby accounting for all of the ICL present, including that around satellite galaxies and projected over all of the cluster galaxies. Additionally, they do not impose \textit{a priori} assumptions on the physical properties of the ICL (e.g., surface brightness, density or morphology).

(v) The Wavelet Decomposition (WD) method, similar to MGF, separates ICL from all galaxies in the cluster \citep[e.g.,][]{DaRocha2005,Guennou2012,Ellien2019, Ellien2021} using a multi-scale approach, where the ICL is usually identified with the lowest frequency component. Similar to the MGF methods, WD also separates galaxies and ICL for the whole image, thereby accounting for all of the ICL present.

Other methods to measure the ICL include the kinematic distribution of planetary nebulae or globular clusters \citep[e.g.,][]{Arnaboldi1996, Alamo-Martinez2017, Hartke2017,Madrid2018,Powalka2018, Harris2020, Hartke2022, Kluge2023}, stacking integral field spectroscopic observations \citep[]{Edwards2016,Edwards2020} as well as image stacking \citep[e.g.,][]{Zibetti2005,Zhang2019, SampaioSantos2021,chen22,Golden-Marx2023}. However, these methods are not considered in this work because kinematic studies are only applicable to a few nearby clusters, integral field spectroscopic observations are currently limited in sample size and stacking analyses are only just starting to provide information on the scaling relationships with their host clusters \citep{Zhang2023}.

Given the difficulty of separating the ICL from the BCG, and other satellite galaxies in the cluster, some works instead measure the fraction of light held by the combination of the BCG and the ICL together, arguing that it is not possible to accurately separate the two (or more) components \citep[BCG+ICL; e.g.,][]{Gonzalez2007, DeMaio2018, Furnell2021, Morishita2017, Kluge2021, Presotto2014, Montes2021, Spavone2020, Zhang2019, SampaioSantos2021}. 

When comparing the observed measurements that have been applied to date, they show significant scatter \citep[e.g.,][]{Montes2022}. It is unclear whether this is physical in origin or due to observational differences (depth, photometric band, measurement method) that are contributing to the scatter. For example, when using SB to measure the ICL, there is an observed trend of increasing ICL fraction with decreasing redshift \citep[e.g.,][]{Burke2015, Montes2022}. However, the CM method shows little evolution at $z < 0.6$ \citep{Montes2022}. \cite{Kluge2021} compared several ways to separate BCG+ICL in their sample of 170 low redshift clusters: using a surface brightness threshold, a luminosity threshold, a double Sérsic decomposition and the excess light above a de Vaucouleurs profile. They find mean ICL fractions that vary from $10$ to $20$ per cent depending on the method used and a mean BCG+ICL fraction of 28 per cent. 

New, deep, wide-field surveys that will increase the samples available for the study of ICL by several orders of magnitude are imminent, e.g. The Vera C. Rubin Observatory's Legacy Survey of Space and Time \citep[LSST; e.g.][]{Robertson2019, Montes2019, Brough2020} and the European Space Agency's Euclid Wide Survey \citep{Euclid2022,EuclidCol2022}. These promise to deliver the large samples needed to explore the ICL as a function of cluster mass, redshift and dynamical state. However, without a detailed analysis of the method by which observers and simulators measure ICL, its interpretation will remain ambiguous. 

Cosmological hydrodynamical simulations are ideal laboratories to explore and isolate the physical mechanisms that form the ICL. They can access the 6D information of each resolution element in the cluster. However, isolating the ICL in simulations is also a complex problem.  In simulations, the methods for quantifying the contribution of ICL include: Aperture-based measures, identifying the ICL as all star particles in a certain radial range from the cluster centre \citep{Pillepich2018b}; kinematic-based measures, separating the ICL on the basis of a double-Maxwellian fit to particle velocities \citep{Dolag2010,remus17} or Gaussian Mixture Models \citep{Proctor2023}; or using the full distribution of star particles in the 6D phase space \citep{Canas_et_al_2019}. 

Several attempts have been made to address the issue of how to define the ICL. \cite{Rudick2011} used a suite of N-body simulations of 6 galaxy clusters ($0.8<M_{\odot} \times 10^{14}<6.5$) specifically tailored to studying ICL \citep{Rudick06} to measure the quantity of ICL found using a number of different methods from the literature (Binding Energy, \citealt{Willman2004, Murante2007, Dolag2010}; Threshold Density, \citealt{Rudick2009} and Surface Brightness threshold, \citealt{Feldmeier2004,Mihos2005, Rudick2010}). They found that techniques that define the ICL solely based on the current position of the cluster luminosity, such as a surface brightness or local density threshold, tend to find less ICL than methods utilizing time or velocity information, including stellar particles' density history or binding energy. They also found that separating the BCG from the surrounding ICL component was a challenge for all ICL techniques, and the differences in the measured ICL quantity between techniques were largely a consequence of the separation of the ICL light projected over the BCG. \cite{Rudick2011} measured a range of ICL fractions across all the clusters using any definition between 9–36 per cent, and within a single cluster different methods changed the measured ICL fraction by up to a factor of two. 

\cite{Cui2014} also compared a dynamical BCG+ICL and ICL fraction separation with a surface brightness threshold in cosmological hydrodynamical simulations of 64 galaxy clusters with ($13.5<$Log$_{10}$M$_{500}/M_{\odot}<15.2$). They found that the dynamical method found higher ICL fractions than the SB method (55 per cent compared to 20-30 per cent).

\cite{Tang2018} investigated the limitations of measuring ICL from optical imaging data using hydrodynamical simulations, testing the impact of the limitations optical images are subject to (e.g. image band, pixel size, surface brightness limit, and point spread function size). Here, we do not investigate the effect of varying these parameters and focus only on the question of measurement method.

There have been advances in both simulations and observational techniques since the \cite{Rudick2011} and \cite{Cui2014}  analyses. For example, the \cite{Rudick2011} simulations have a large luminous particle mass of $1.4~\times~10^6 M_{\odot}$ and did not include hydrodynamic evolution and so neglected certain aspects of galaxy and cluster evolution which may play a role in determining the spatial distribution of luminous material in the cluster. This included not being able to resolve galactic cores and so they did not attempt to test composite models on their simulations. On the observational side, new methods based on theoretical data analysis considerations are being developed to provide new flexible approaches to ICL measurements, with more evolved MGF and WD techniques like the CICLE (MGF) and DAWIS (WD) algorithms applied here. 

To better facilitate future ICL investigations with the next-generation of facilities, we have assembled a cross-section of theorists and observers working on this topic to test the robustness and biases associated with different ICL measurement methods. These include theorists working with different simulations and observers who span the range of techniques currently employed for ICL analyses. The aim of the work presented here is to assess the different definitions of ICL in both observations and simulations, to determine their fidelity and enable robust comparisons between observations and simulations. We apply eight currently-used observational BCG+ICL and ICL techniques to mock images of 61 galaxy clusters from four of the most widely used cosmological hydrodynamical simulations (Horizon-AGN, \citealt{Dubois.2014};  Hydrangea, \citealt{Bahe_et_al_2017}; Illustris-TNG, \citealt{Nelson2019}; and Magneticum, \citealt{dolag17}). We then compare the results obtained with the observational methods with the amount of ICL predicted in the simulations. 

The layout of the paper is as follows: Section~\ref{sec:sims_methods} describes the four simulations used in this analysis, the method used to create mock images for the observational analyses and the simulation-based measures of BCG and ICL applied to these simulations. Section~\ref{sec:obstech} describes the eight different observation-based measures of BCG and ICL applied to the mock images. Section~\ref{sec:results} presents our comparison of these different measures. We discuss our results in the context of recent research in Section~\ref{sec:disc} and draw our conclusions in Section~\ref{sec:concl}. Throughout this work, we assume the native cosmology of each of the simulations as described in Section~\ref{sec:sims_methods}.

\section{Simulations and theoretical quantities}
\label{sec:sims_methods}

\subsection{Galaxy clusters from cosmological simulations}
\label{sec:sims}

In this study, we compare the outcome of a diverse range of methods intended to extract ICL properties from observed and simulated clusters of galaxies. We hence apply these methods to simulated clusters from a range of cosmological $\Lambda$CDM hydrodynamical simulations.  By using simulated objects, instead of observed images, we can access all of the information content provided by the underlying simulation data and extract ICL properties as typically measured within the numerical and theoretical community.

We aimed to target relaxed clusters with mass $\gtrsim10^{14}M_\odot$, namely haloes that are massive enough to have significant amounts of ICL, but not so massive that too few would be present in currently-available cosmological simulations. For these, we require sufficiently good numerical resolution so that the diffuse stellar component of the ICL is properly sampled, i.e. with large numbers of stellar particles. We choose and analyze galaxy clusters from four different state-of-the-art hydrodynamical cosmological simulation suites: Magneticum, Horizon-AGN, Hydrangea, and IllustrisTNG. These allow us to perform our ICL-focused comparisons by marginalizing over the possible effects of 1) different numerical methods to solve for the coupled equations of gravity and hydrodynamics, 2) different numerical mass and spatial resolutions, 3) different adopted cosmology assumptions, 4) different halo finders, and, chiefly, 5) different choices and implementations of the underlying galaxy-formation astrophysical models, such as feedback processes. The four different simulation suites have been described and used extensively in the literature over the past few years: we summarize salient aspects of each of them in the following and in Table~\ref{tab:sims}.

\subsubsection{Horizon-AGN}
\label{sec:horizon} 
Horizon-AGN \citep{Dubois.2014} is a cosmological-volume hydrodynamical simulation performed using \textsc{RAMSES} \citep{Teyssier.2002}, an adaptive mesh refinement (AMR)-based Eulerian hydrodynamics code. 
An initial 142 comoving Mpc-length box contains $1024^3$ dark matter particles each with a mass of $8\times10^7~\MSUN$. An initially-uniform $1024^3$ cell gas grid is refined according to a quasi-Lagrangian criterion, with the smallest cell sizes fixed at 1 physical kpc.

The implemented subgrid physics include the following processes: gas cooling via Hydrogen and Helium cooling with a contribution from metals down to $10^{4}$~K \citep{Sutherland.1993}; the star formation is modelled via a Schmidt law with standard 2 per cent efficiency \citep{Kennicutt.1998} and feedback from Type II, Type Ia supernovae and stellar winds. Black holes include a high-efficiency \emph{quasar mode} with isotropic injection of thermal energy and a low-efficiency \emph{radio mode} with cylindrical bipolar outflows and jet velocity of $10^{4}~{\rm km\,s^{-1}}$ following \citet{Omma.2004}. The stellar particles, i.e. the resolution elements that constitute the ICL, have a mass resolution of about $2\times10^6\,\MSUN$.

\subsubsection{Hydrangea}
\label{sec:hydrangea} 
Hydrangea \citep[see also \citealt{Barnes_et_al_2017}]{Bahe_et_al_2017} is a suite of 24 cosmological hydrodynamical zoom-in simulations of massive galaxy clusters using a variant of the EAGLE simulation mode \citep{Schaye.2015}. Similar to Magneticum, the simulations are based on the SPH code \textsc{gadget-3} \citep{springel05}. Sub-grid models are used for gas cooling, star formation, the associated mass and energy feedback, as well as the growth of and feedback from supermassive black holes. For details on their implementation, we refer the interested reader to \cite{Schaye.2015} and \cite{Bahe_et_al_2017}, but note here that particular care was taken to calibrate the efficiency of supernova and black hole feedback to observations of stellar masses and sizes, as well as the gas content of group-scale haloes.

As demonstrated by \citet{Bahe_et_al_2017} and \citet{Ahad_et_al_2021}, the predicted stellar mass function of satellite galaxies matches observations closely both in the local Universe and out to at least $z \approx 1.5$. The total stellar mass within $z \approx 0$ clusters is also realistic, although the BCGs are too massive by a factor of 2--3 compared to observations \citep{Bahe_et_al_2017}. The latter is not unique to Hydrangea; it is likely that it reflects shortcomings in the AGN feedback model that also lead to overly high gas fractions and central entropy cores as discussed by \citet[see also \citealt{Oppenheimer2021}]{Barnes_et_al_2017}. We note that the substructure identification used in Hydrangea includes an additional step that removes stars bound to satellites more rigorously than the standard \textsc{Subfind} algorithm (Bah\'{e} et al., in preparation) and therefore, tends to lead to a lower mass of stars associated with the BCG and ICL.

\subsubsection{IllustrisTNG}
\label{sec:tng}
The Next Generation Illustris\footnote{\url{www.tng-project.org}} (IllustrisTNG) is a suite of cosmological magnetohydrodynamical (MHD) simulations of galaxies of three different comoving volumes each performed at varying resolution levels. The flagship runs of the series are called TNG100, TNG300 and TNG50, and in this paper we make exclusive use of the TNG100 run \citep{Pillepich2018b, Nelson2018, Springel2018, Marinacci2018,Naiman2018, Nelson2019}. Tens of thousands of galaxies are therein evolved across a period-boundary box of 110 comoving Mpc a side and with stellar/gas particle resolution of $1.4\times10^6\,\MSUN$, i.e. mass resolution similar to that of Hydrangea and Horizon-AGN (Table~\ref{tab:sims}).

In contrast to the other simulation suites of this paper, IllustrisTNG includes MHD. It is based on a moving-mesh code, \textsc{AREPO} \citep{Springel2010}, which combines the benefits of both grid (as in Horizon-AGN) and lagrangian (as in Magneticum and Hydrangea) codes. 

Similar to the other simulation models, the IllustrisTNG simulations, and hence TNG100, are the results of a rich ensemble of coupled astrophysical processes acting across spatial and time scales, including star formation, gas cooling and heating, stellar evolution and metal enrichment, feedback from stars and seeding, growth and feedback from SMBHs. The details of the IllustrisTNG model are described by \citet{Weinberger2017} and \citet{Pillepich2018a}, and succinctly summarized and compared to the other suites in Table~\ref{tab:sims}.

There are 280, 14, and 2 clusters more massive than $10^{14}\,\MSUN$ in the TNG300, TNG100, and TNG50 volumes at $z=0$, respectively. Their stellar mass content and their BCG and satellite populations have been extensively characterized and compared to observations by \citet{Pillepich2018b} and by \citet{Joshi2020} in terms of morphological transformations, \citet{Pulsoni2020, Pulsoni2021} for their stellar kinematics, \citet{Donnari2021} in terms of quenched fractions. Of particular relevance for this work, \citet{Ardila2021} had shown, with an apples-to-apples comparison to deep Hyper Suprime-Cam (HSC) observations, that the outer stellar masses of TNG100 galaxies in $\sim10^{14}\,\MSUN$ haloes are consistent with weak-lensing inferences to better than 0.12 dex.

\subsubsection{Magneticum}
\label{sec:magneticum}
Magneticum Pathfinder\footnote{\url{www.magneticum.org}} is a suite of fully hydrodynamical cosmological simulations covering a large range in simulation volumes and resolutions. All simulations were performed with an updated version of the TreePM-Smoothed-Particle Hydrodynamics (SPH) code GADGET-3 based on GADGET-2 \citep{springel05}. They also include updates to the SPH formulation with respect to the treatment of viscosity \citep{dolag05,beck16}, the SPH kernels \citep{donnert13,beck16}, and the thermal conductivity \citep{dolag04}. The implemented subgrid physics contains supermassive black hole (SMBH) treatment and Active Galactic Nuclei (AGN) feedback \citep{fabjan10,hirschmann14}, star formation and metal enrichment from Supernovae Ia, Supernovae II and Asymptotic Giant Branch stars according to \citet{tornatore04,tornatore07}, as well as cooling processes coupled to the local metallicity following \citet{wiersma09,dolag17}. Kinetic feedback from stellar winds is included according to \citet{springel03}. 

In this paper, we include galaxy clusters from two of the simulation volumes of the Magneticum suite: \textit{Box2b} at the high resolution (HR) level, and \textit{Box4} at the ultra-high resolution (UHR) level. \textit{Box4} is a volume of $(68~\mathrm{comoving \, Mpc})^3$, with initially $2\times576^{3}$ particles at the UHR resolution level. The individual mass resolution is $\sim 2.6\times10^{6}\,\MSUN$ for stellar particles, with their gravitational softening being $\sim 1$~kpc. \textit{Box2b} has a volume of $(909~\mathrm{comoving \, Mpc})^3$, with $2\times2880^{3}$ particles at the HR resolution level: this corresponds to a mass resolution of $\sim 5\times10^{7}\, \MSUN$ for stellar particles and gravitational softening for the stellar component of $\sim2.8$~kpc. See papers above for more details on the numerical resolution of all matter components. 

Note that for the Magneticum simulations, one gas particle can spawn up to four stellar particles, and thus the stellar particle mass quoted here is just the average stellar mass and can be substantially smaller than the initial gas particle mass. Both box volumes have been used to study galaxy and galaxy cluster properties in prior works, most notably for the study presented here are those on the ICL and BCG properties \citep{remus17}, early cluster and BCG formation \citep{remus22b} and stellar halo properties \citep{remus22}, galaxy populations in galaxy clusters \citep{lotz19}, and substructure properties \citep{kimmig22}, as well as the general introductory papers on halo-to-stellar mass properties \citep{teklu17} and AGN properties \citep[e.g.,][]{hirschmann14}.

\subsection{Selection of simulated clusters}
\label{sec:select_clust}
From the simulations described above, we select galaxy clusters at $z = 0$ in the halo mass range $\log_{10}\,(M_\mathrm{200c}\,/\MSUN{}) = [14.0, 14.5]$, whereby $M_\mathrm{200c}$ denotes the mass enclosed within a spherical overdensity of 200 times the critical density. 

As the Magneticum \textit{Box4} simulation covers a small volume, it harbours only three galaxy clusters with masses larger than $M_\mathrm{crit}\geq1\times10^{14}M_\odot$, of which only one is relaxed as preferred for this study. The much larger \textit{Box2b}, on the other hand, realizes more than 1000 clusters, from which we select 13 clusters with low total substructure masses, as this is a good indicator for relaxed galaxy clusters \citep[e.g.,][]{kimmig22}. We only explicitly apply a relaxedness criterion to the Magneticum systems and discuss the effects of this choice in Section~\ref{sec:disc-relax}. We have indicated the Box4 cluster separately in the figures introducing the different simulations, to show that its properties are consistent with those of the other simulations which have a similar box size (Table~\ref{tab:sims}).

These cuts resulted in a final sample of 61 simulated clusters, with 9, 27, 11, and 14 clusters from Horizon-AGN, Hydrangea, TNG100 and the two boxes of Magneticum respectively.  Of the 61 clusters, 29 are relaxed by visual inspection. We analyze this sample throughout the following sections.

\subsection{Finding structures and substructures}

To identify galaxies and satellite galaxies within the large cosmological simulated volumes, and thus to isolate the BCG and the ICL, haloes and subhaloes need to be located. For the Magneticum, Hydrangea and IllustrisTNG runs, we use the output of the simulations based on the baryonic version of the \textsc{Subfind} halo finder \citep[see also \citealt{Springel_et_al_2001}]{dolag09} to identify gravitationally-bound (sub)structures. The versions of these halo finders used on the three aforementioned projects are not identical, but are very similar. In contrast, Horizon-AGN, uses the \textsc{AdaptaHOP} halo finder \citep{Tweed2009}. 

The \textsc{Subfind} and \textsc{AdaptaHOP} codes differ in terms of how they define particle membership to (sub)haloes:

\textsc{Subfind} identifies substructures that are both locally overdense and gravitationally bound. In the initial step, haloes are identified through a Friends-of-Friends (FoF) algorithm. This is run on the dark matter particles only, with baryon particles assigned to the FoF halo (if any) of their nearest DM neighbour. Within each  FoF halo, substructures are then identified by searching for local density peaks, now considering all types of resolution elements and particles. Different subhaloes are separated by saddle points in the density field, with each subhalo limited to particles within the isodensity contour passing through its limiting saddle point. An iterative unbinding procedure is then applied to each subhalo to remove any particle/cell that is not gravitationally bound to it. Finally, all resolution elements not assigned to a substructure are considered as members of the central subhalo, after applying the same iterative unbinding process. This procedure is based on the kinetic (and for gas, internal) energy of each particle, and as such is not directly comparable to observationally feasible approaches. A noteworthy limitation of this approach is that by design any resolution element that lies beyond the limiting isodensity contour is ignored, even if it is in fact gravitationally bound to the subhalo (e.g.~\citealt{Muldrew_et_al_2011, Canas_et_al_2019}).

\textsc{AdaptaHOP} is a fully topological code that does not feature an unbinding procedure. Particles are first sorted into groups around peaks in the density field that are linked to other groups at saddle points. Each structure is then hierarchically divided into smaller groups in steps of increasing density. Haloes are defined as a group-of-groups linked by saddle points that exceed 160 times the mean dark matter density and groups within each halo are hierarchically regrouped so that each substructure has a smaller mass than the host (sub)structure. The absence of an unbinding procedure implies that different numbers of particles and resolution elements are associated to structures and substructures by \textsc{AdaptaHOP} in comparison to \textsc{Subfind}, and hence to haloes vs. subhaloes and galaxies vs. satellites.

These differences between (sub)halo finders are non-trivial. For example, different subhalo finders will clearly leave an impact on what it means to ``excise'' the contribution of satellite galaxies from the mass of the BCG and of the ICL \citep[e.g.][]{Knebe2011}. However, they encompass what is typically done in the field by different research groups and thus provide us with yet another opportunity to account for possible systematic differences. Moreover, the removal of the light/mass from satellites is also performed in a variety of ways observationally (Section~\ref{sec:obstech}). We hence proceed as is and comment on possible differences below.

Finally, in the case of the Magneticum, Hydrangea and IllustrisTNG simulations, the total cluster masses, defined throughout this paper based on $M_\mathrm{200,crit}$, do not depend on the functioning of the \textsc{Subfind} and FoF algorithms. Namely, once a FoF halo and its centre are identified, the latter being the deepest point of the potential well, spherical-overdensity masses are measured accounting for {\it all} particles and resolution elements in the volume, irrespective of whether they belong to the FoF or \textsc{Subfind} structure. In the case of Horizon-AGN, the $M_\mathrm{200,crit}$ masses are based on the particles and resolution elements that are deemed by \textsc{AdaptaHOP} to belong to a given halo. Based on the cluster centres found as described above, we extract cubes around each halo with a side length of $4\mathrm{~Mpc}$ that are used to generate the mock observations (Section~\ref{sec:mocks}).

\subsection{Idealized, Simulation-based measures of BCG and ICL}
\label{sec:simtech}

For all galaxy clusters, we define a radius of $1\mathrm{Mpc}$ around the central galaxy (the BCG), comparable to the cluster virial radius at these cluster masses. The true total stellar mass within this sphere, including all satellite galaxies, is denoted $M_\mathrm{*,Tot}$. Furthermore, the stellar mass within this sphere that is not allocated to a satellite galaxy comprises both the BCG and the ICL, and we refer to this component as $M_\mathrm{BCG+ICL}$. We calculate the mass fraction of the BCG+ICL with respect to the whole stellar mass within this sphere as $f_\mathrm{BCG+ICL} = M_\mathrm{BCG+ICL}/M_\mathrm{*,Tot}$. The (simulated) mass fraction is different from the (observed) luminosity fraction and depends on how the mass-to-light ratio differs between the ICL and the galaxy populations. We explore this further in Section~\ref{sec:disc-cons}.

Separating the ICL from the BCG is more complicated than separating the substructures from the main body in simulations (as well as in observations), as a simple binding criterion is not sufficient to achieve this. Therefore, in this work, we compare two different methods which are applied to separate those components in simulations. These are described in the following. 

\subsubsection{Aperture-based measures}
A simple and robust method to define the ICL in simulations is to consider all star particles in a certain radial range from the cluster centre. In this approach, the (fixed) inner cut is used to separate the ICL from the BCG. As discussed by \citet{Pillepich2018b}, the choice of this radius ($r_\mathrm{inner}$) is somewhat ad-hoc, although commonly used observational definitions of the BCG extent (e.g.~Petrosian or Kron radii) typically correspond to around 30--100~kpc. We therefore separately calculate ICL fractions with $r_\mathrm{inner}$ = {30, 50, 100}~kpc in this study. These are 3D radii, i.e., spheres so the simulated ICL measurements are not made in projection. In each case, we only consider star particles associated with the main halo of the cluster, i.e.~with all satellites excised\footnote{This definition excludes ``fuzz'' particles that are completely unbound from the cluster, but we have verified that such particles contribute $\ll$ 1 per cent to the ICL.}.

\subsubsection{Kinematic-based measures}
The stellar light of a galaxy cluster, after subtracting the substructures, has been shown to consist of two kinematically distinct components \citep[e.g.][]{Dressler79,Nelson2002,bender:2015,longobardi:2015}. These two components have been found in simulated galaxy clusters as well, and have been associated to the inner BCG and the outer diffuse stellar component, or ICL \citep{Dolag2010,remus17}. The velocity component of a galaxy cluster can be described by a double-Maxwellian distribution in 3D space, which in projection resembles a double-Gaussian distribution \citep[see][for more details]{remus17}. Unfortunately, separating the ICL and BCG through this kinematic measure often does not resemble the radial separation found if a double-S\'ersic profile is fit to the radial density distribution of the ICL and BGC component \citep{remus17}, indicating that the kinematic and spatial measures trace different aspects of the ICL and BCG and highlighting the need to define these two components self-consistently.

\subsection{Mock images of simulated clusters}
\label{sec:mocks}
One of the most robust methods for the comparison of simulations with observational data is through the analysis of synthetic `mock' observations \citep[e.g.][]{Jonsson2006,Naab2014,Choi2018,Baes2020,Olsen2021}, which enable us to measure quantities in the same way as we would observationally. In making these synthetic observations, we consider future idealized LSST-like images created using the method described in \citet{Martin2022}. We summarise how we produce mock images for each of the clusters in our sample below. 

Mock images are produced for each cluster by extracting all star particles in a (4~Mpc)$^{3}$ cube centred around each BCG. The spectral energy distribution (SED) for each star particle is then calculated, based on its age and metallicity, from a grid of \citet{Bruzual2003} simple stellar population models assuming a \citet[][]{Chabrier2003} IMF.\footnote{As noted in \citet{Martin2022}, different IMFs have close to equal effect on the brightness of the BCG and the ICL component so that the only qualitative impact on our results would be to increase or decrease the overall surface brightness of the image.} Unlike \citet{Martin2022}, we choose to neglect the effect of dust attenuation on the SED of each particle due to the different stellar evolution recipes, feedback schemes and hydrodynamics codes employed by each simulation. This can have a strong effect on the diffusion and distribution of metals or dust and therefore the amount of attenuation. Additionally, since we focus on the ICL where very little dust should be present, modelling dust attenuation is only relevant for observational predictions for the flux of the member galaxies. The luminosity of each star particle is calculated by first summing the resultant luminosity of the attenuated SEDs once they have been redshifted to $z=0.05$ and convolved with the LSST $r$-band transmission functions \citep{Olivier2008}.

We employ an adaptive smoothing scheme in order to better represent the distribution of stellar mass in phase space and remove unrealistic variations between adjacent pixels.\footnote{The adaptive smoothing code used in this paper is available from \href{https://github.com/garrethmartin/smooth3d}{github.com/garrethmartin/smooth3d}} We follow a similar procedure to the \textsc{adaptiveBox} method employed by \citet[][]{Merritt2020}, by splitting each particle into 500 smaller particles which are then re-distributed in 3-D according to a Gaussian distribution centred on the position of the original particle and with standard deviation set by the distance to the original particle's 5th nearest neighbour.

Finally, a 2-D image is created by collapsing the particles along one of the axes and summing the flux across a 2D grid with elements of $0.2^{\prime\prime}\times0.2^{\prime\prime}$. For every cluster, we produce smoothed mock images in 3 projections ($xy$, $xz$ and $yz$). Each image is convolved with a point spread function (PSF)\footnote{We use the $g$-band Hyper Suprime-Cam \citep[HSC;][]{Miyazaki2012} 2D PSF measured by \citet[][]{Montes2021} to $289^{\prime\prime}$ and extrapolated to $420^{\prime\prime}$ based on a power law fit. The PSF FWHM is always larger than the smoothing length in regions of interest (i.e. for the clusters).} and random Gaussian noise is added to simulate a predicted LSST 10-year limiting surface brightness of $\mu_r=$30.3 mag/arcsec$^{2}$ (P. Yoachim, private communication).

There is no variation in the noise level across the image and also choose to neglect other instrumental and astrophysical contaminants (e.g., foreground and background objects, Galactic cirrus, scattered light, ghosts and diffraction spikes) which may be present in real imaging. Our results, therefore, represent a best-case scenario for the various methods presented in this paper.

\begin{figure*}
	\includegraphics[width=0.75\textwidth]{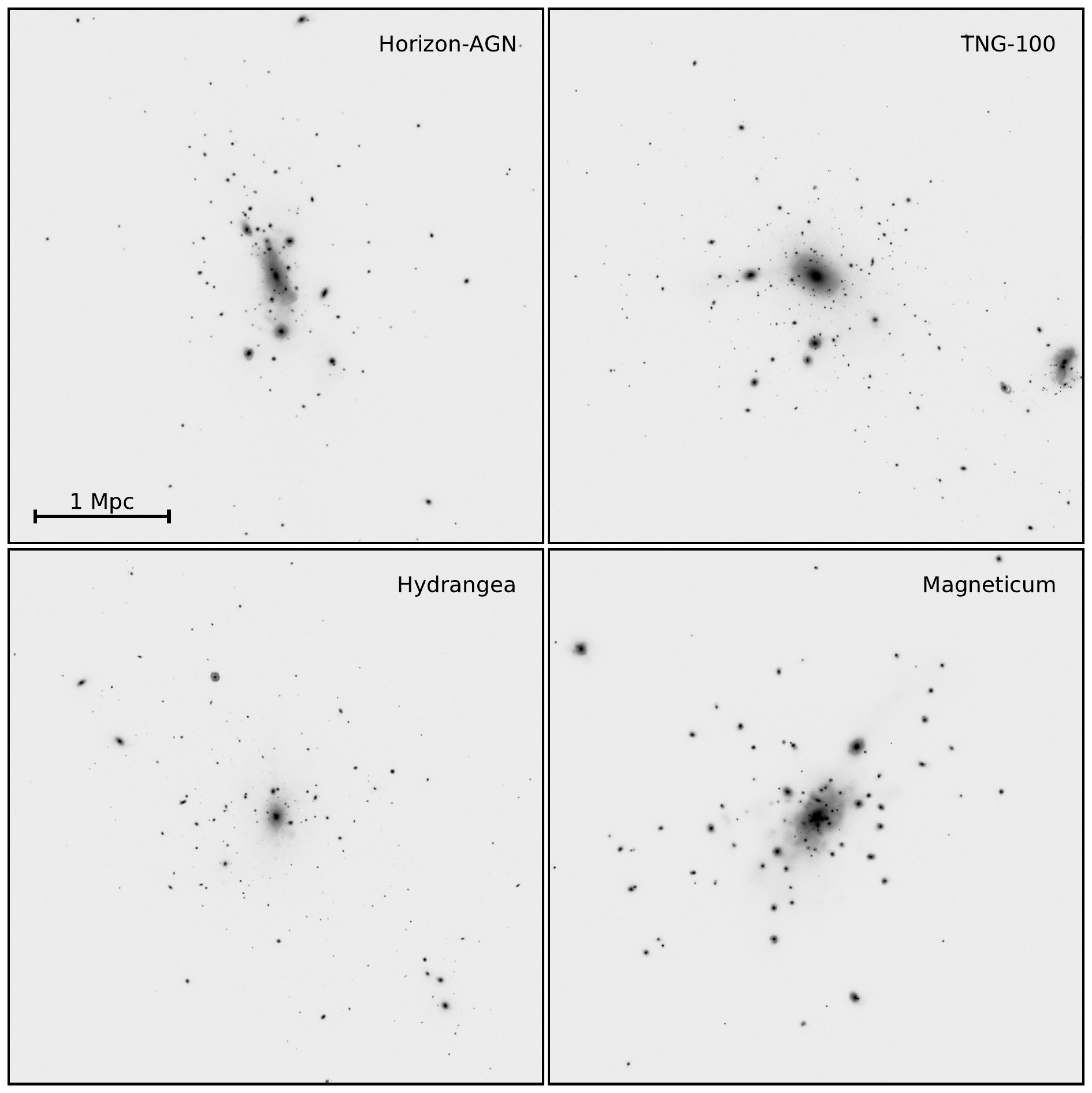}
    \caption{Log-scaled mock images of a random relaxed cluster from each simulation. Gaussian noise is added to each image to simulate a limiting surface brightness $\mu_r=30.3$ mag/arcsec$^{2}$.}
    \label{fig:demo_images}
\end{figure*}

Fig.~\ref{fig:demo_images} shows an example of an $r$-band smoothed mock image for one cluster from each simulation. In these images the lightest shade corresponds to a surface brightness fainter than $30.3$~mag/arcsec$^{2}$. 

\section{Observational Techniques}
\label{sec:obstech}
Deep images of clusters of galaxies show that the transition between the BCG and the ICL happens smoothly without a clear break point. Therefore, observers have had to devise techniques to study these components either together (BCG+ICL) or to separate them in order to study the ICL separately. In this Section we describe the eight observational algorithms to measure the BCG+ICL and/or ICL fraction of total cluster luminosity considered in this paper. These are presented grouped by the type of parent method: Surface Brightness Threshold in Section~\ref{sec:sblim}, Non-Parametric Measures in Section~\ref{sec:nonpara}, Composite Models in Section~\ref{sec:profilefits}, Multi-Galaxy Fitting in Section~\ref{sec:multi} and Wavelet Analysis in Section~\ref{sec:wavelet}. Each of these methods is carried out by different people, each of whom applies different pre-processing steps before they make the measurements. Therefore, in this work we are testing complete image processing and analysis methods, not only different ICL methods, to determine how well the different groups' measurements compare to one another.

In order to calculate the BCG+ICL and ICL fractions as a function of the total cluster luminosity, the total luminosity of each cluster is measured by summing the luminosity in a circular aperture of radius $R=1$~Mpc centered on the BCG. This outer radius was set to remove cluster radius as a potential source of uncertainty in the fraction measures.

\subsection{Surface brightness threshold (SB)}
\label{sec:sblim}

The easiest approach to separating the ICL from the galaxies in the cluster from an observational point of view is to use a surface brightness threshold. This method defines all light below a certain surface brightness threshold as the ICL. The method accounts for the contribution to the ICL from the outskirts of any of the cluster galaxies instead of only the BCG. Observations and simulations have shown that this method does a reasonable job in separating the BCG and the diffuse light \citep[e.g.,][]{Feldmeier2004, Rudick2011, Cui2014} and that there are physical arguments for a surface brightness threshold of $\mu_V = 26.5$ mag/arcsec$^2$. However, using this definition in observations is more complicated as the different surface brightness depths of different images lead to different amounts of ICL being measured. It also misses the ICL projected over any of the galaxies in the cluster. 

In this work, we have adopted a surface  brightness threshold of $\mu_r > 26$ mag/arcsec$^2$. This method is denoted `SB Martinez-Lombilla' and `Montes' hereafter depending on the observer making the measurement \citep[e.g.][]{Montes2021, Martinez-Lombilla2023a}. The ICL contribution is then the sum of the flux in all the pixels fainter than this threshold value and brighter than the surface brightness limit of the mock images ($\mu_r = 30.3$ mag/arcsec$^2$). Those pixels are within a circular aperture (Martinez-Lombilla)  or elliptical (Montes) of $R \sim 1$~Mpc centered on the BCG. The ellipticity of the aperture in the Montes method is based on the ellipticity of the BCG at large radius. The Martinez-Lombilla method applied a 2x2 binning to the images (i.e. reduced the image size by 4, so the spatial resolution is 0.4$^{\prime\prime}$/pixel) to ensure the analysis code ran in a reasonable time.

\subsection{Non-Parametric Measures (NP)}
\label{sec:nonpara}

\subsubsection{1D non-parametric extraction}
In this method, denoted `Gonzalez' hereafter, we follow the approach of \citet{DeMaio2018}, which builds on \citet{Gonzalez2007}, in which the BCG+ICL surface brightness is extracted in a series of logarithmically-spaced circular annuli.  As in \citet{DeMaio2018}, the first step for this approach is to mask all detected galaxies in the image other than the BCG.\footnote{In a few cases two merging BCGs are each left unmasked.}  Cluster galaxies, which are detected with \textsc{SExtractor}, are masked with elliptical apertures extending out to three times the Kron radius. For each cluster a few galaxies that lie close to the centroid of the BCG, and hence are not detected by \textsc{SExtractor}, are manually masked.

Next, the median surface brightness is calculated within logarithmically-spaced annuli of width $d\log r=0.15$. The sky level is taken to be the median pixel value at $r>1.9$~Mpc and this sky level is subtracted from the profile level.  While the simulations contain no sky contribution, this step was included to mimic true observations. The total flux within 1~Mpc is then calculated by integrating the surface brightness in apertures extending out to this radius, using 1D interpolation to match the radial boundaries. 

To calculate the fractions, we sum the \texttt{AUTO} fluxes from \textsc{SExtractor} for all of the galaxies detected within the same 1~Mpc radius and take the ratio of these two fluxes. This approach makes no assumptions regarding the shape of the ICL profile, but because it relies on the median within annular apertures it may underestimate the total ICL if there are strong tidal features that are not well reflected in the median values. 

\subsubsection{2D non-parametric extraction}
\label{sec:selmask}
In this method, denoted `Martinez-Lombilla' hereafter, the BGC+ICL is directly measured from the mock images following the procedures described in \cite{Martinez-Lombilla2023a}. This method consists of constructing a mask in which every source is masked, including faint tidal tails of any kind, only allowing the BCG and ICL flux to remain. The mask is built from Python scripts using a threshold for detections of 1.1~$\sigma$ above the image background level. Due to the wide variety of objects in the field, we use the ``hot+cold'' masking method \citep[e.g.][]{Rix2004, Montes2021, Martinez-Lombilla2023a, Martinez-Lombilla2023b}. As overlapping sources are frequently found in galaxy clusters (i.e. galaxy cluster members that overlap the ICL and the BCG), we unsharp-masked the original image prior to the application of the hot mask to increase the contrast. To unsharp-mask, we convolved the image with a Gaussian filter \citep[e.g.][]{Montes2021, Martinez-Lombilla2023a, Martinez-Lombilla2023b} with $\sigma=5$~pixels, and subtracted it from the original image. Finally, we radially increased all the masks as required by visual identification to avoid including any source of faint light from the outskirts of the satellite galaxies in our BCG+ICL measurements. Then, we measure the BCG+ICL flux by summing the flux of the masked images within a circular aperture of $R \sim 1$~Mpc around the BCG. We applied 2x2 binning to the images to speed up the analysis code.

\subsection{Composite models (CM)}
\label{sec:profilefits}
The stellar envelope in the outer part of BCGs is observed to be an additional component to the single or double empirical model profiles (often Sérsic) that reproduce the inner regions of massive galaxies, as noted in several works \citep[e.g.][]{Seigar2007, Gonzalez2007, Donzelli2011, Iodice2016, Spavone2017b, RAGUSA2021, Ragusa2022FrASS...952810R}. The methods in this section include composites of multiple analytic models for light distribution of the BCG+ICL components. These methods account for the ICL projected over the BCG, but will fail to capture any component of ICL that is not symmetrically centered on the BCG.

\subsubsection{1D de Vaucouleurs profile decomposition}

In this method, denoted `Ahad' hereafter, we measure the fraction of light in the ICL component compared to the total cluster light within 1~Mpc radius using a single de Vaucouleurs profile fitting method as described in detail in \citet{Ahad2023}. We first mask all the galaxies in the mock image except for the BCG by running \textsc{SExtractor} \citep{Bertin1996}. The \textsc{SExtractor} segmentation maps are radially extended by 40~kpc before creating the masks to ensure that most parts of the diffuse light in the outskirts of satellite galaxies are excluded in our measurement. Then we measure the azimuthally averaged BCG+ICL surface brightness profiles in logarithmic circular apertures centred on the BCG and fit the BCG light using a de Vaucouleurs profile. The BCG profile is then subtracted from the BCG+ICL profile to obtain the excess light at the outskirts, which we identify as the ICL and integrate out to the surface brightness limit of the mock image (or 1~Mpc, whichever is smaller) to measure the total light in ICL. The total light in the BCG+ICL is measured by integrating the BCG+ICL surface brightness profile out to the same surface brightness limit as was used for the ICL, stated above. 

The fraction of light in the BCG+ICL and ICL is obtained by dividing the total light in the corresponding components by the total cluster light within 1~Mpc radius, including the BCG, satellite galaxies, and the ICL component. 

\subsubsection{1D multi-component decomposition}

In this method, denoted `Ragusa' hereafter, we derive the total contribution of the faint outskirts of the BCG (stellar envelope plus ICL) as the integrated light from the transition radius ($R_{\rm{tr}}$) outwards by performing a 1D multi-component decomposition of the BCG azimuthally-averaged surface brightness profiles, using 2 S{\'e}rsic profiles as described in detail in \citet[]{RAGUSA2021,Ragusa2022FrASS...952810R,ragusa2023}. $R_{\rm{tr}}$ is
the distance from the galaxy centre where the contribution from
the galaxy outskirts (i.e., stellar envelope plus diffuse light) starts
to dominate the total light distribution. We model and subtract the BCGs in 2D (to their $R_{\rm{tr}}$) from the mock images. We then carefully mask all the sources in the residual image (for the mock images these are just the cluster satellite galaxies) and then measure the ICL luminosity by fitting an exponential law to reproduce the diffuse ICL component and sum all the pixels beyond the transition radius.

In order to derive the ICL fraction we measure the total cluster luminosity by summing the contributions of all the satellite galaxies, the BCG up to its $R_{\rm{tr}}$ and the ICL component. We also derived the BCG+ICL fraction, which is the luminosity of the ICL component plus that of the BCG up to its $R_{\rm{tr}}$. Although the mock images do not have a contribution from the observed sky, the added noise must be taken into account given the low surface brightness of the ICL. In studying observational data, it is crucial to avoid edge effects in estimating residual background fluctuations.  We estimate the average value of the background fluctuations by fitting the light in circular annuli of constant steps of 30~kpc between $r=1.7-1.9$~Mpc, centered on the centre of the cluster, having carefully masked all the satellite galaxies. This average value, and its rms, are taken into account in all of the estimated values.

\subsection{Multi-galaxy fitting (MGF)}
\label{sec:multi}
The Multi-Galaxy Fitting methods model and remove all the galaxies in the image with either traditional analytical profiles \citep[e.g.][]{Giallongo2014, Morishita2017, Poliakov2021} or orthonormal mathematical bases \citep{Jimenez-Teja2016, Jimenez-Teja2018}. These methods separate galaxies and ICL for the whole image, thereby accounting for all of the ICL present, including that projected over galaxies and around satellite galaxies.

\subsubsection{CICLE}
\label{sec:cicle}

CHEFs Intracluster Light Estimator \citep[CICLE,][]{Jimenez-Teja2016,Jimenez-Teja2018, Jimenez-Teja2019,Jimenez-Teja2021,jimenez-teja2023,deOliveira2022,Dupke2022} is an algorithm that creates two-dimensional models of the galaxies to disentangle them from the ICL. All galaxies are detected with \textsc{SExtractor} and fit using orthonormal mathematical bases composed by Chebyshev rational functions and Fourier series \citep[CHEFs,][]{Jimenez-Teja2012}. The use of orthonormal bases guarantees that all morphologies -- independently of the level of substructure, asymmetry, or irregularity -- can be fit by the linear composition of the elements of the basis. The fact that Chebyshev polynomials do not tend to zero at the infinite end makes it possible to recover all the light from the extended wings of the galaxies. Additionally, Chebyshev polynomials are optimal to interpolate functions in their domain of definition, a property that is directly inherited by the CHEF bases. This means that CHEF models are built using a small number of components (typically, 10 Chebyshev rational functions and 10 Fourier modes) and a higher number of elements is only needed if the galaxy is very large or shows a great level of detail. CHEF models are computed down to the noise level of the image or until the stellar haloes of the galaxies converge asymptotically, so it is straightforward to build models for all satellite galaxies in the cluster. However, for the particular case of the BCG (and its extended halo, if it is present) CHEFs will model the galaxy and the ICL together, due to the spatial coincidence of the peak of the two surfaces in projection. Then, the limits of the BCG-dominated region are defined prior to the modelling, using a change in the curvature (the difference in the slope of the BCG+ICL composite surface) as the criterion to disentangle the BCG from the ICL. The fit is made in two-dimensions and does not make any prior assumption on the shape or possible symmetry of the ICL or the BCG. We obtain an ICL map by removing all CHEF models of the galaxies. If we just re-add the CHEF model of the BCG, we obtain the BCG+ICL map, with all satellite galaxies excised. Final ICL and BCG+ICL fractions are measured using these maps, estimating the flux within the fixed 1Mpc-radius aperture used in this work.  This method is denoted `CICLE' hereafter.

The CICLE method applied a 2x2 binning to the images to speed up the processing.

\subsection{Wavelet Decomposition}
\label{sec:wavelet}
The Wavelet Decomposition method separates ICL from all galaxies in the cluster using a multi-scale approach. Like Multi-Galaxy Fitting this method also separates galaxies and ICL for the whole image, thereby accounting for all of the ICL present.

\subsubsection{DAWIS}
\label{sec:dawis}

\texttt{DAWIS} \citep[Detection Algorithm with Wavelets for Intracluster light Studies;][]{Ellien2021} is a recent addition to a series of multiscale, wavelet-based algorithms optimized for low surface brightness astronomy \citep{Adami2005,DaRocha2005,DaRocha2008, Ellien2019}. Such algorithms use wavelet representation \citep{Slezak1994,Starck2007} and multi-resolution vision models \citep{bijaoui1995} to i) disentangle the signal associated with small details from large scale variations in analyzed images ii) model the noise and detect sources down to very faint surface brightness iii) model the 2D light distribution of these sources. The novelty of \texttt{DAWIS} compared to previous wavelet-based algorithms is its iterative approach: it only models a few sources at once, starting with the brightest, and removes them from the image. It then repeats the process until it converges on a residual map containing only noise. 

The sources detected and modelled in each iteration usually do not correspond to entire astrophysical objects, but rather to substructures. The information content is dissected into small pieces, denoted atoms. Since no astronomical prior is given to the algorithm, the nature of one atom alone is purely artificial, and relies on how \texttt{DAWIS} estimates and captures significant signal in the wavelet space at each iteration. However, it is possible by selecting these atoms to synthesize images of actual astrophysical objects. The most trivial synthesis is the sum of all atoms of an image, which provides a completely de-noised version of the astrophysical field.

To select atoms, three properties of interest are: the wavelet scale $z$ at which it has been detected by \texttt{DAWIS}, the size $S$ of the detected atom, and the spatial position of its intensity maximum in the image. Different classification schemes are tested utilizing these three parameters:

i) The hard wavelet threshold method is denoted `DAWIS-W' hereafter. This separates based on the wavelet scale of atoms alone, without any other prior. The idea behind such a criterion is that a wavelet transform is a series of convolutions with a dilated kernel of size $2^z$ pixels. Therefore, each wavelet scale $z$ corresponds (roughly) to a characteristic size $2^z$. It is assumed here that the characteristic extent of the ICL in astronomical images is much larger than the characteristic size of galaxies. Therefore, the atoms associated with galaxies are expected to be detected mainly at small wavelet scales, while the atoms associated with the ICL are expected to be detected mainly at large wavelet scales. A hard separation can be performed by setting a specific wavelet scale as threshold \citep[an approach taken by][]{Ellien2021}. In this work the threshold is set to the wavelet scale z = 6.  Within this scheme, the BCG is treated similarly to the rest of the satellite galaxies, and the atoms are classified either as `galaxy' or as `ICL'. Including spatial information as an extra step allows atoms to be classified as either `galaxy' or `BCG+ICL'. This is done by inserting a constraint for atoms classified as galaxies, which must be outside a radius $r_\mathrm{BCG}$ from the centre of the image (corresponding to the centre of the BCG). 

ii) This size separation method is denoted `DAWIS-SS' hereafter and uses the size of restored atoms as a separation criterion rather than the wavelet scale. While both approaches appear similar, they provide different results. This is due to the fact that the actual size of detected atoms does not increase linearly with the wavelet scale. In this scheme, atoms are classified either as `galaxy' or `ICL', and the BCG is also treated similarly to satellite galaxies, or, by including spatial information, atoms are classified into `galaxy' and `BCG+ICL'. The atom size threshold used in this work to separate ICL from galaxies is 150 kpc.

iii) The mixture of a wavelet-based analysis and the surface brightness threshold method (Section~\ref{sec:sblim}) is denoted `DAWIS-SB' hereafter. All the atoms of the image are summed to synthesize the entire de-noised galaxy cluster field, to which the surface brightness threshold is then applied. The main difference with the regular surface brightness threshold is that all sources have been detected through wavelet analysis leading to different limiting depths. No separation is made between the BCG and the rest of the satellites for this method.

The three schemes tested for this analysis are of an over-simplistic nature as they are based on arbitrary single-value criteria. It is unlikely that they correctly capture all of the morphological differences of a whole cluster sample. However, this analysis provides a first glance of the performance of DAWIS and shows how it compares to other measurements. While more complex selections are possible, they are beyond the scope of this study.

Note that in order to reduce computation time, the cluster images analysed by \texttt{DAWIS} were rebinned by a factor of 4.

\section{Results}
\label{sec:results}
Here we consider the BCG+ICL and ICL fractions measured directly from the simulations and using the observers' methods from the mock images. 

\subsection{Simulated BCG mass, BCG+ICL and ICL fractions}
\label{sec:theory_res}
\begin{figure}
    \includegraphics[width=\columnwidth]{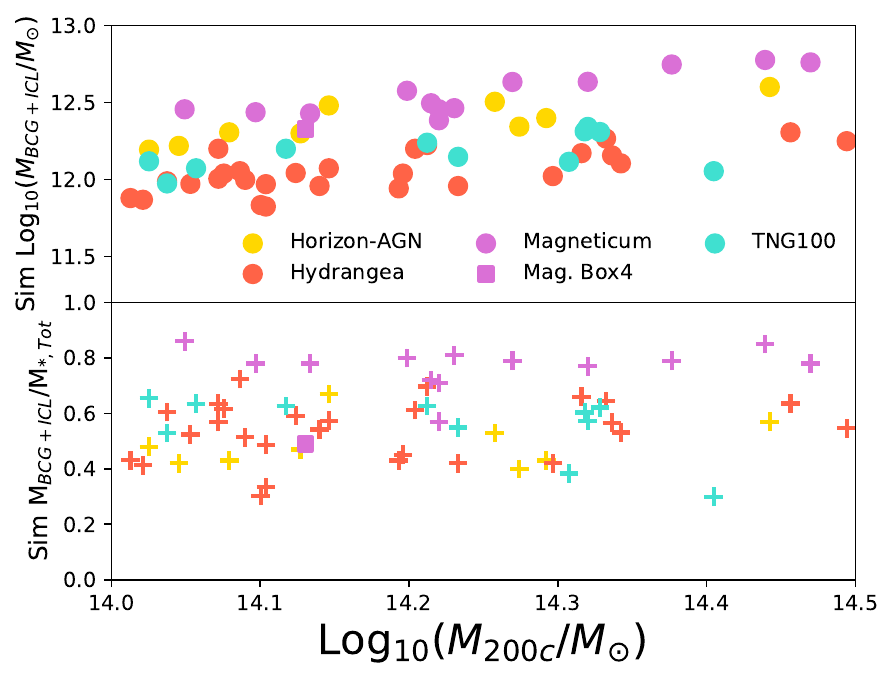} 
    \includegraphics[width=\columnwidth]{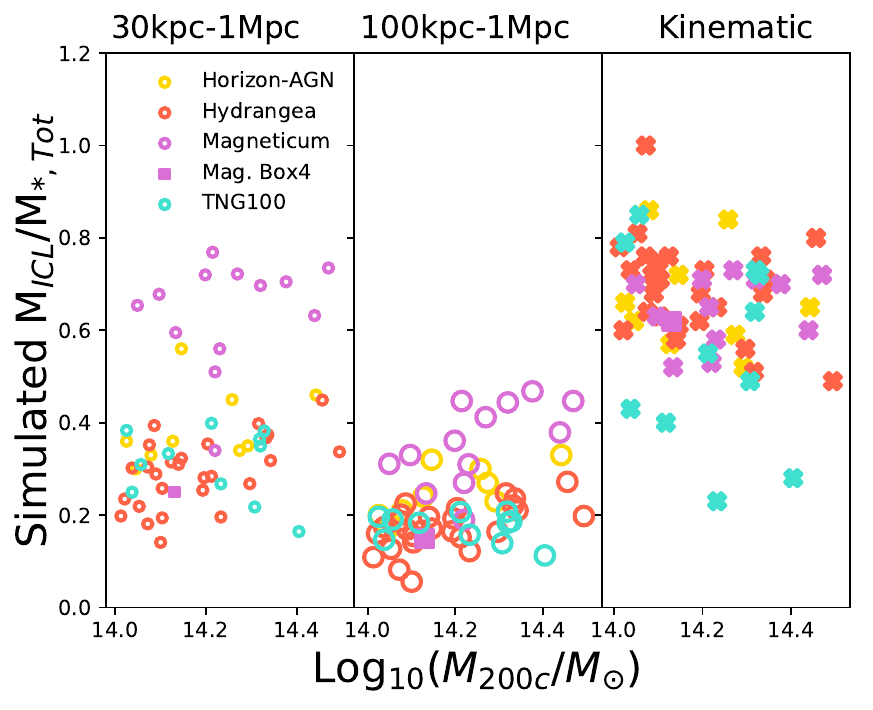} 
    \caption{The simulated BCG+ICL mass (upper panel) and the simulated BCG+ICL fraction (middle panel) as a function of cluster mass. The lower panel shows the ICL fraction measured from the simulations in 2 different apertures (left panel: 30~kpc$-$1~Mpc; middle panel: 100~kpc$-$1~Mpc) and the right-hand panel shows the ICL fraction measured using kinematic separation (crosses) as a function of cluster mass. The different simulations are indicated by the legend and include the one cluster from the higher-resolution Magneticum Box4 simulation.
    }
    \label{fig:Sims_bcgmass}
\end{figure}

The upper panel of Fig.~\ref{fig:Sims_bcgmass} shows the simulated BCG+ICL mass compared to the cluster mass for each of the 61 simulated clusters across the 4 simulations. The middle panel of Fig.~\ref{fig:Sims_bcgmass} shows the BCG+ICL fractions, i.e. $M_{\rm{(BCG+ICL)}}/M_{*,\rm{Tot}}$, measured directly from the simulations in an aperture of radius $0-1$~Mpc. The simulated BCG+ICL mass increases with increasing cluster mass in the top panel, as expected given the underlying BCG-halo mass relationship \citep[e.g.][]{Brough2008,Lidman2012}, but the simulated BCG+ICL fraction does not increase with cluster mass suggesting that the satellite galaxy contribution also increases over this cluster mass range. The simulated BCG+ICL fractions are given in Table~\ref{tab:sim_lum_mass} and range from $0.49\pm0.08$ for Horizon-AGN to $0.75\pm0.10$ for Magneticum.  Throughout we give the $1\sigma$ standard deviation as the scatter around these mean values.

The lower panel of Fig.~\ref{fig:Sims_bcgmass} shows the ICL fractions, i.e. $M_{\rm{(ICL)}}/M_{*,\rm{Tot}}$, measured directly from the simulations with three different methods indicated in the panel. The left-hand panels show two different aperture measures (with radii 30~kpc$-$1~Mpc and 100~kpc$-$1~Mpc; for conciseness we do not show the 50~kpc$-$1~Mpc aperture measures) and the right-hand panel shows the kinematic separation, as a function of cluster mass coloured by the four different simulations. As would be expected, we observe that the ICL fraction varies as a function of the aperture that it is measured within, with ICL fraction decreasing as the aperture range decreases from 30~kpc$-$1~Mpc to 100~kpc$-$1~Mpc. The simulated ICL fractions are given in Table~\ref{tab:sim_lum_mass} and fall from $0.38\pm0.16$ for the 30~kpc$-$1~Mpc aperture to $0.22\pm0.09$ for the 100~kpc$-$1~Mpc aperture. The lower panel of Fig.~\ref{fig:Sims_bcgmass} also shows that the kinematic method of separating ICL measures a higher ICL fraction with a mean ICL fraction of $0.65\pm0.13$. This includes a Hydrangea cluster with a kinematic ICL fraction of $1.0$ owing to a massive starburst in its BCG. We do not observe a relationship of BCG+ICL or ICL fraction with host cluster mass across this mass range.

Fig.~\ref{fig:Sims_bcgmass} and Table~\ref{tab:sim_lum_mass} show that the Magneticum clusters have a higher BCG+ICL mass and BCG+ICL and aperture ICL fractions than the other simulations. This is a result of the selection of very-relaxed systems from this simulation, most of the selected clusters are at the uppermost range of BCG+ICL fraction compared to the full Magneticum cluster sample. The middle panel of Fig.~\ref{fig:Sims_bcgmass} shows that the BCG+ICL fraction for the higher resolution Magneticum Box4 simulation can be seen to be consistent with the fractions for the other three simulations. We discuss this further in Section~\ref{sec:disc-relax}. With this exception, we do not observe any further substantial differences in the BCG+ICL mass, BCG+ICL or ICL fractions between the four different simulations. Given the non-trivial differences between the two halo finders used, this suggests that the relevant quantities are calculated robustly and means that we can proceed in our analysis considering the simulations as a whole. 

\begin{table*}
\caption{Data for the different simulations. Mean BCG+ICL fractions. Mean ICL fractions over 3 of the simulated measures. Mean (Observed-Simulated fractions) for each of the simulations: BCG+ICL is for the $0-1$~Mpc aperture and ICL is for the $100$~kpc$-1$~Mpc aperture. Comparing simulated BCG+ICL fractions (0 - 1~Mpc aperture) and ICL fractions (100~kpc - 1~Mpc aperture) measured in mass compared to luminosity, i.e. $F_{\rm{M}}/F_{\rm{L}}$: ($M_{\rm{BCG+ICL}}$/M$_{*,\rm{Tot}}$)/(L$_{\rm{BCG+ICL}}$/L$_{\rm{Tot}}$ or ($M_{\rm{ICL}}$/M$_{*,\rm{Tot}}$)/(L$_{\rm{ICL}}$/L$_{\rm{Tot}}$.}
\label{tab:sim_lum_mass}
\begin{tabular}{l c c c c c c c c c}
\hline
Simulation&BCG+ICL Fraction &ICL Fraction&ICL Fraction&ICL Fraction& Mean (Obs-Sim) &Mean (Obs-Sim)& $F_{\rm{M}}/F_{\rm{L}}$ & $F_{\rm{M}}/F_{\rm{L}}$\\
&$0-1$~Mpc&30~kpc-1~Mpc&100~kpc-1Mpc&Kinematic&BCG+ICL&ICL&$0-1$~Mpc&100~kpc-1~Mpc\\
\hline
Horizon-AGN & 0.49 $\pm$ 0.08 & 0.39 $\pm$ 0.08 & 0.25 $\pm$ 0.05 & 0.67 $\pm$ 0.11 & -0.05 $\pm$ 0.07 & -0.08 $\pm$ 0.05 & 1.10 $\pm$ 0.29 & 1.39 $\pm$ 0.34 \\
Hydrangea & 0.54 $\pm$ 0.11 & 0.29 $\pm$ 0.07 & 0.17 $\pm$ 0.05 & 0.69 $\pm$ 0.10 & -0.06 $\pm$ 0.10 & -0.06 $\pm$ 0.05 & 0.99 $\pm$ 0.10 & 0.98 $\pm$ 0.62 \\
Magneticum & 0.75 $\pm$ 0.10 & 0.61 $\pm$ 0.15 & 0.34 $\pm$ 0.10 & 0.64 $\pm$ 0.07 & -0.07 $\pm$ 0.07 & -0.17 $\pm$ 0.10 & 0.96 $\pm$ 0.04 & 1.10 $\pm$ 0.16 \\
TNG100 & 0.55 $\pm$ 0.11 & 0.31 $\pm$ 0.07 & 0.17 $\pm$ 0.03 & 0.56 $\pm$ 0.20 & -0.03 $\pm$ 0.08 & -0.05 $\pm$ 0.04 & 1.09 $\pm$ 0.10 & 1.18 $\pm$ 0.47 \\
\hline
Overall Mean & 0.58 $\pm$ 0.14 & 0.38 $\pm$ 0.16 & 0.22 $\pm$ 0.09 & 0.65 $\pm$ 0.13 \\
\hline
\end{tabular}
\end{table*}

\subsection{Observational BCG+ICL Analyses}
\label{sec:ICLBCG}
\begin{figure}
	\includegraphics[width=\columnwidth]{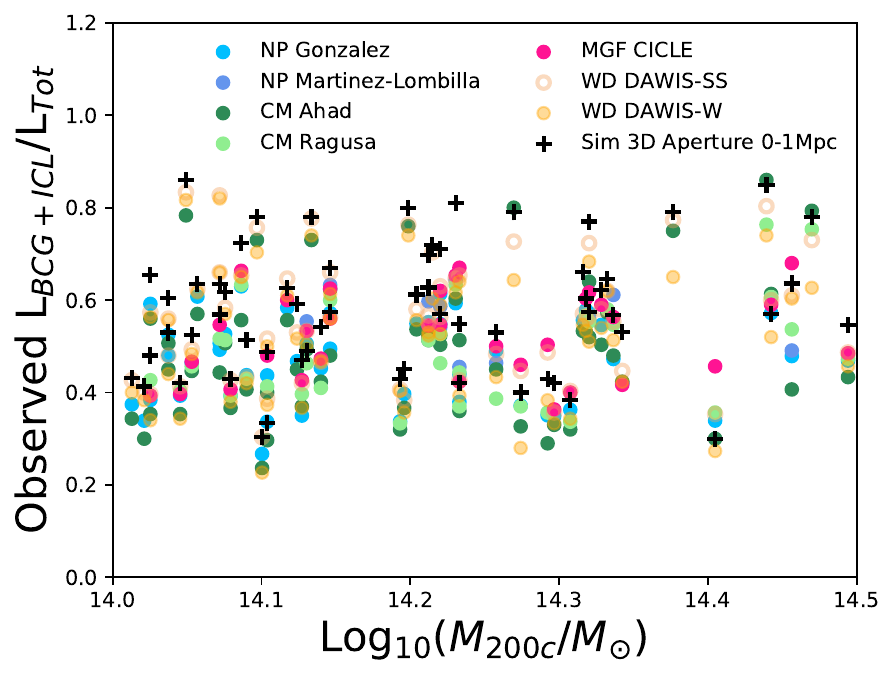}
	\includegraphics[width=\columnwidth]{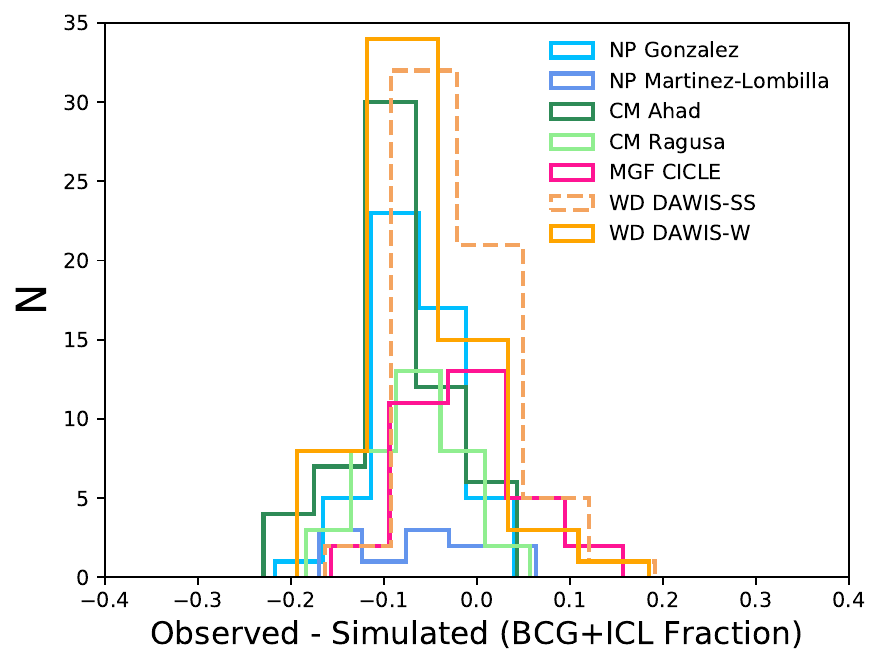}
    \caption{
    Observed BCG+ICL fraction (mean measurement over the measured projections). The upper panel shows the observed BCG+ICL fraction as a function of cluster halo mass coloured by measurement type for the 61 simulated clusters across the 4 simulations. We do not observe a dependence of any of the measures on cluster halo mass. The lower panel shows the difference between the Observed BCG+ICL fraction and the Simulated 0 - 1~Mpc Aperture measurement. The observed measurements are presented grouped by measurement type: Non-Parametric Measures (NP; Gonzalez and Martinez-Lombilla), Composite Models (CM; Ahad and Ragusa), Multi-Galaxy Fitting (MGF; CICLE) and Wavelet Decomposition (WD; DAWIS-SS and DAWIS-W). The Surface Brightness Threshold method is not included for BCG+ICL fractions as it removes the BCG by definition. The numbers of clusters measured are different for each of the observed measures.  This figure demonstrates that all methods agree to <0.1 dex in excising the contribution of satellites.}
    \label{fig:ICLBCG_Frac_clustermass}
\end{figure}

In the upper panel of Fig.~\ref{fig:ICLBCG_Frac_clustermass} we present the BCG+ICL fractions measured by the observers' methods from the 2D mock images and in the lower panel a comparison between those observed measures and the BCG+ICL fraction measured directly from the 3D simulations in the $0-1$~Mpc aperture, which by definition includes the BCG. Each observer has measured the BCG+ICL fraction in at least one of the three projections of the simulations ($xy, xz, zy$). In these plots we present the mean over those projections (L$_{\rm{BCG+ICL}}$/L$_{\rm{Tot}}$) and will consider in Section~\ref{sec:proj} the scatter in the measurements as a result of projection effects. The observed measurements are presented grouped by measurement type: Non-Parametric Measures (Gonzalez and Martinez-Lombilla), Composite Models (Ahad and Ragusa), Multi-Galaxy Fitting (CICLE) and Wavelet Decomposition (DAWIS-SS and DAWIS-W). The Surface Brightness Threshold method is not included for BCG+ICL fractions as it removes the BCG by definition. Table~\ref{tab:BCGICL_observer_results} gives the numbers of clusters measured for each of the observed measures. This is different for each of the measures due to different levels of manual intervention being required and observer availability to undertake that. The mean BCG+ICL fractions are also given in  Table~\ref{tab:BCGICL_observer_results} and range from $0.47\pm0.09$ for Gonzalez to $0.56\pm0.06$ for Martinez-Lombilla and $0.56\pm0.12$ for DAWIS-SS with an overall mean BCG+ICL fraction of $0.51\pm0.12$. We do not observe a dependence of any of the measures on cluster halo mass in this narrow mass range. 

The lower panel of Fig.~\ref{fig:ICLBCG_Frac_clustermass} shows the histogram of the difference between the observers' BCG+ICL fractions and the simulated 0-1 Mpc fractions. We find that the observed BCG+ICL fractions are generally slightly lower than the simulated measurements. The means of these differences are given in Table~\ref{tab:BCGICL_observer_results} and range from $-0.02\pm0.12$ for DAWIS-SS and $-0.02\pm0.06$ for CICLE to $-0.08\pm0.05$ for Ahad. The overall Mean (Observed - Simulated) $=-0.05\pm0.09$.  

For some measurements, more light is found by the observed methods than by the simulated method, i.e., a higher BCG+ICL fraction. These numbers are given in Table~\ref{tab:BCGICL_observer_results} and can be seen to occur more frequently for the non-parametric measure of Martinez-Lombilla $(N(>0)/N_{\rm{tot}})=0.27$, multi-galaxy fitting CICLE $(N(>0)/N_{\rm{tot}})=0.30$, and the wavelet decomposition measures of DAWIS-W $(N(>0)/N_{\rm{tot}})=0.28$ and DAWIS-SS $(N(>0)/N_{\rm{tot}})=0.34$. 

The DAWIS measures also show the largest standard deviation compared to the simulated measure of all the observational measures, equivalent to a fractional uncertainty of 12 per cent.

We explored whether the mean Observed-Simulated differences depend on the simulation the clusters are sourced from. The mean differences are given in Table~\ref{tab:sim_lum_mass} and are consistent within the standard deviations, ranging from $-0.03\pm 0.08$ for TNG100 to $-0.07\pm 0.07$ for Magneticum.

\begin{table*}
\caption{Data for different observational methods of measuring BCG+ICL fractions. $N_{\rm{tot}}$ gives the numbers of clusters measured for each of the observed measures. Mean BCG+ICL fraction is the mean fraction over all the clusters measured by that observer. Mean (Observed - Simulated) is the mean difference between the observed BCG+ICL fractions and the simulated 0 - 1~Mpc aperture measure. $N>0$ gives the number of clusters with an Obs - Sim difference > 0. Mean Projection Scatter quantifies projection effects and is described in Section~\ref{sec:disc}. The uncertainties are the $1\sigma$ standard deviations.}
\label{tab:BCGICL_observer_results}
\begin{tabular}{l c c c c c}
\hline
Observer & $N_{\rm{tot}}$ & Mean BCG+ICL &Mean (Obs - Sim) & N>0 & Mean Projection \\
 & & fraction& & & Scatter \\
 \hline
\textit{Non-Parametric Measure}\\
Gonzalez & 51 & 0.47 $\pm$ 0.09 & -0.07 $\pm$ 0.05 & 4 & 0.06 $\pm$ 0.04 \\
Martinez-Lombilla & 11 & 0.56 $\pm$ 0.06 & -0.05 $\pm$ 0.08 & 3 & 0.07 $\pm$ 0.05 \\
\hline
Mean & & 0.49 $\pm$ 0.09 & -0.07 $\pm$ 0.05 && 0.06 $\pm$ 0.04 \\
\hline
\textit{Composite Model}\\
Ahad & 59 & 0.49 $\pm$ 0.15 & -0.08 $\pm$ 0.05 & 6 & 0.05 $\pm$ 0.04 \\
Ragusa & 34 & 0.49 $\pm$ 0.11 & -0.07 $\pm$ 0.05 & 2 & 0.03 $\pm$ 0.02 \\
\hline
Mean & & 0.49 $\pm$ 0.13 & -0.08 $\pm$ 0.05 && 0.04 $\pm$ 0.03 \\
\hline
\textit{Multi-Galaxy Fitting}\\
CICLE & 33 & 0.52 $\pm$ 0.09 & -0.02 $\pm$ 0.06 & 10 & 0.06 $\pm$ 0.04 \\
\hline
\textit{Wavelet Decomposition}\\
DAWIS-SS & 61 & 0.56 $\pm$ 0.12 & -0.02 $\pm$ 0.12 & 21 & 0.11 $\pm$ 0.08 \\
DAWIS-W & 61 & 0.52 $\pm$ 0.12 & -0.06 $\pm$ 0.12 & 17 & 0.14 $\pm$ 0.11 \\
\hline
Mean& & 0.54 $\pm$ 0.12 & -0.04 $\pm$ 0.12 && 0.13 $\pm$ 0.10 \\
\hline
\hline
Overall Mean & & 0.51 $\pm$ 0.12 & -0.05 $\pm$ 0.09 && 0.08 $\pm$ 0.08 \\
\hline
\end{tabular}
\end{table*}

\subsection{Observational ICL Analyses}
\label{sec:ICL}
\begin{figure*}
	\includegraphics[width=0.9\textwidth]{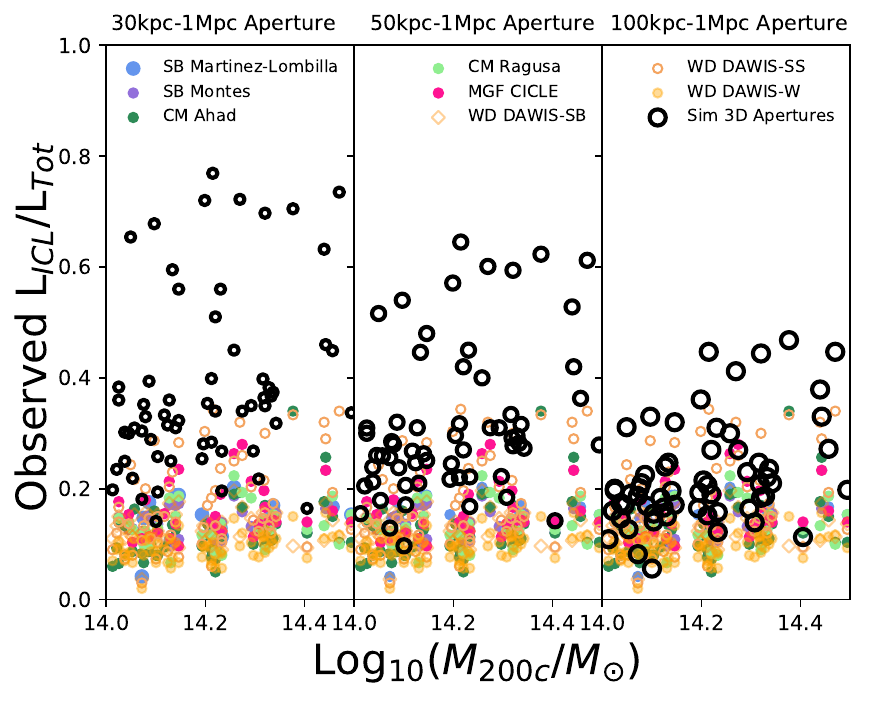}
    \caption{Observed ICL fraction (mean measurement over simulated projections) as a function of cluster halo mass coloured by measurement type for the 61 simulated clusters across the 4 simulations. The different panels show the three Simulated aperture measures ($30-1$~Mpc, $50-1$~Mpc, $100-1$~Mpc). We do not observe a dependence of any of the measures on cluster halo mass.}
    \label{fig:ICL_Frac_clustermass_aper}
\end{figure*}

\begin{figure}
	\includegraphics[width=\columnwidth]{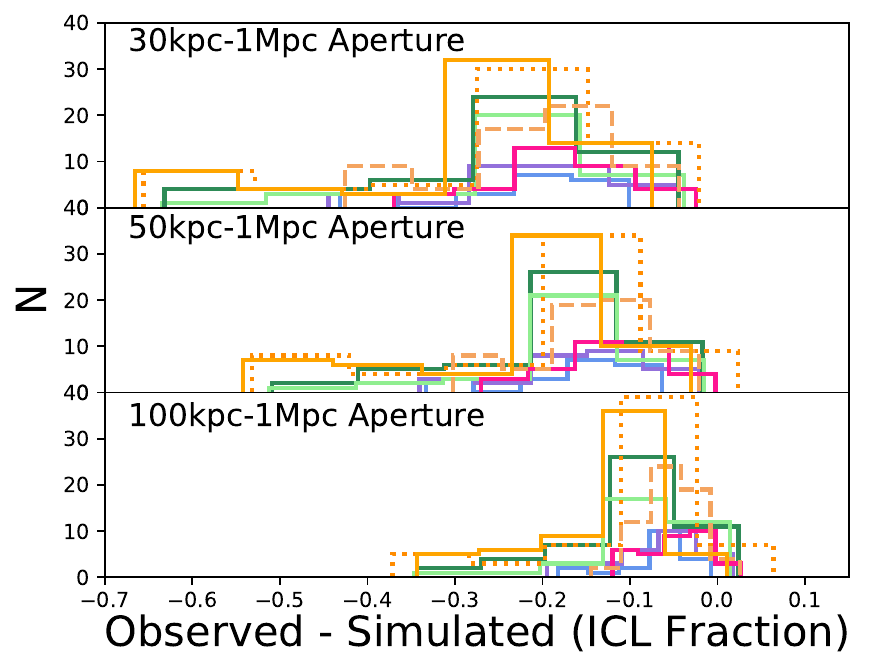}
    \caption{The difference between the observed ICL fraction and the simulated measurements for the $30-1$~Mpc, $50-1$~Mpc and $100-1$~Mpc apertures. The colours for the different histograms are the same as given in Figure~\ref{fig:ICL_Frac_clustermass_aper}.}
    \label{fig:ICL_Frac_ap_histo}
\end{figure}

While BCG+ICL fractions are challenging to measure, subtracting the BCG to estimate the ICL fraction alone, i.e., (L$_{\rm{ICL}}$/L$_{\rm{Tot}}$), is even more challenging. In this Section we present the ICL fractions measured by the observers from the mock images. Figs~\ref{fig:ICL_Frac_clustermass_aper} and ~\ref{fig:ICL_Frac_ap_histo} present the observed ICL fractions compared to the three simulated aperture fractions, which subtract a 30, 50 or 100~kpc of the inner radius of the cluster. Again, we present the mean over the measured projections and consider in Section~\ref{sec:proj} the scatter in the measurements as a result of projection effects. The observed measurements are presented grouped by measurement type: for ICL fractions these include Surface Brightness threshold (SB Martinez-Lombilla and Montes), Composite Models (Ahad and Ragusa), Multi-Galaxy Fitting (CICLE) and Wavelet Decomposition (DAWIS-SB, DAWIS-SS and DAWIS-W).  The mean ICL fractions are given in Table~\ref{tab:ICL_observer_results}. They are lower than the BCG+ICL fractions and range from $0.09\pm0.02$ for DAWIS-W to $0.17\pm0.08$ for DAWIS-SS with an overall mean ICL fraction of $0.13\pm0.05$. Table~\ref{tab:ICL_observer_results} also gives the numbers of clusters measured for each of the observed measures. This is different for each of the measures due to different levels of manual intervention required and observer availability to undertake that. We do not observe a dependence of any of the ICL fraction measures on cluster halo mass in this narrow cluster mass range. 

Fig~\ref{fig:ICL_Frac_ap_histo} shows that the observed ICL fractions are generally lower than the simulated aperture measurements. The Observed-Simulated difference is largest for the 30~kpc$-1$~Mpc aperture and decreases moving to the 100~kpc$-1$~Mpc aperture. The overall mean Observed-Simulated differences are given in Table~\ref{tab:ICL_observer_results} and range from $-0.24\pm0.13$ for the 30~kpc$-1$~Mpc aperture to $-0.09\pm0.08$ for the 100~kpc$-1$~Mpc aperture. Given that the simulated 100~kpc$-1$~Mpc aperture ICL fraction is the closest to the observed ICL fractions we adopt this as the fiducial simulated ICL fraction hereafter. The closest observational measures are CICLE and DAWIS-SS ($-0.05\pm0.04$ and $-0.05\pm0.06$) and the most discrepant is the DAWIS-W method ($-0.13\pm0.09$). The mean Observed-Simulated differences are slightly larger than those for the BCG+ICL separation but have a similar scatter  ($0.09\pm0.08$). Table~\ref{tab:ICL_observer_results} also gives the number of clusters with a mean Observed - Simulated difference $>0$. There are fewer measures than for the BCG+ICL fractions, due to the excision of the BCG light. The DAWIS-SS wavelet decomposition technique finds the most cases with $N(>0)/N_{\rm{tot}}=0.18$ with all other methods finding fractions of $N(>0)/N_{\rm{tot}}=0.03-0.11$. 

Table~\ref{tab:ICL_observer_results} includes the two Surface Brightness Threshold measures (SB Martinez-Lombilla and Montes). The methods for these measures are the most similar in this analysis and the identical mean and scatter found for the 100~kpc$-1$~Mpc aperture (i.e. Observed-Simulated $= -0.07\pm0.05$) suggests that small differences from one observer to the next (in this case circular apertures and binning the data, SB Martinez-Lombilla vs elliptical apertures and not binning, Montes) do not have a significant impact.

We also explored whether the mean Observed-Simulated differences depend on the simulation the clusters are sourced from. The mean differences are given in Table~\ref{tab:sim_lum_mass} and are consistent within the standard deviations, with the exception of Magneticum which has a larger offset $-0.17\pm0.10$.

\begin{figure}
	\includegraphics[width=\columnwidth]{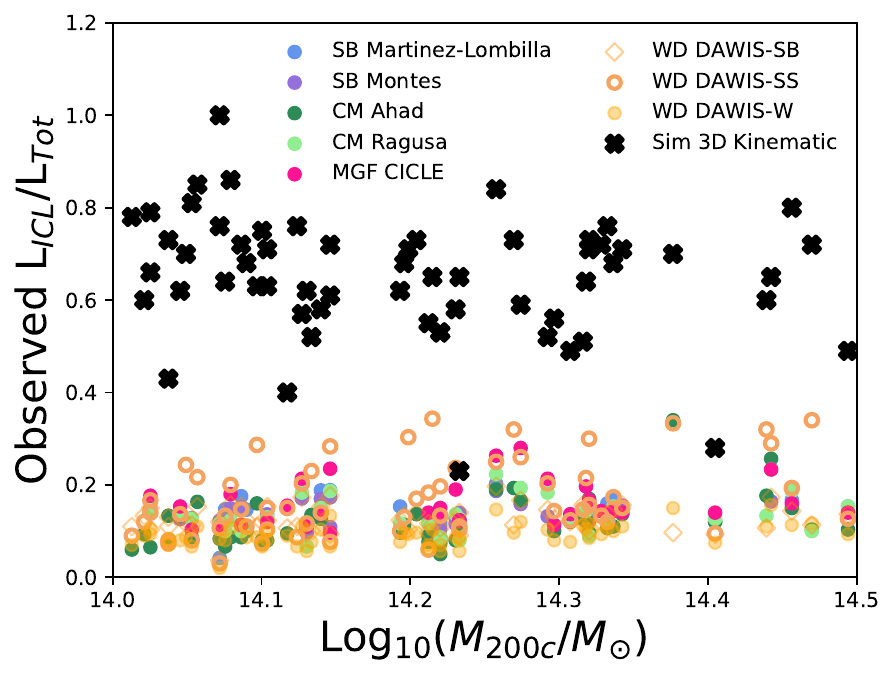}
	\includegraphics[width=\columnwidth]{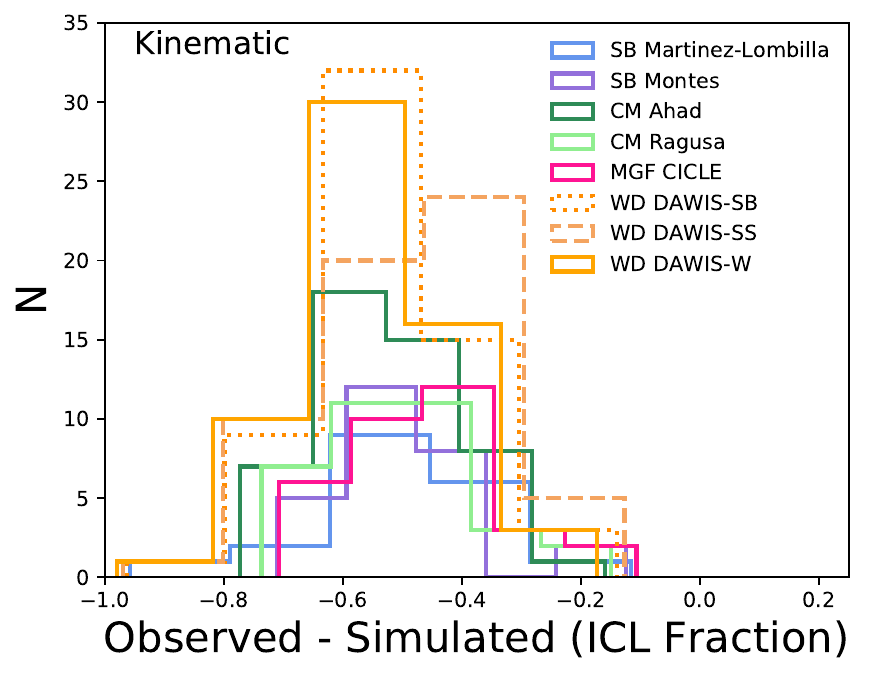}
    \caption{The upper panel shows the observed ICL fraction (mean over measured simulation projections) as a function of cluster halo mass shown with the kinematic measures made on the simulations. The lower panel shows the difference between the Observed ICL fraction and the Simulated kinematic measurement. The numbers of clusters measured are different for each of the observed measures.  This figure demonstrates that the observed methods are offset by 0.5 dex with respect to the kinematic measures.}
    \label{fig:ICL_Frac_clustermass_kin}
\end{figure}

In the upper panel of Fig~\ref{fig:ICL_Frac_clustermass_kin} we present the ICL fractions measured by the observers compared to the ICL fraction measured through the kinematic method from the simulations. The lower panel shows the difference between those observed measures and the simulated kinematic fractions. The simulated ICL fractions measured with the kinematic method are significantly larger than the observed fractions and the simulated aperture fractions shown in Fig~\ref{fig:ICL_Frac_clustermass_aper}. The mean differences are given in Table~\ref{tab:ICL_observer_results} and range from $0.46\pm0.14$ for CICLE to $0.56\pm0.13$ for DAWIS-W. Overall, the mean (Observed-Simulated) $=-0.51\pm0.14$. The mean Observed-Simulated differences are significantly larger than for the aperture methods and the scatter around those means is also significantly larger.

\begin{table*}
\caption{Data for different observational methods of measuring ICL fraction. $N_{\rm{tot}}$ gives the number of clusters measured for each of the observed measures. Mean ICL fractions for each of the observed methods. Mean (Observed - Simulated) is the mean difference between the observed ICL fractions and the 4 simulated measures (Apertures: 30~kpc - 1~Mpc, 50~kpc - 1~Mpc, 100~kpc - 1~Mpc; Kinematic). $N>0$ gives the number of clusters with an Obs - Sim (100~kpc - 1~Mpc Aperture) difference $>0$. Mean Projection Scatter quantifies projection effects and is described in Section~\ref{sec:disc}. The uncertainties are the $1\sigma$ standard deviations around the mean values.}
\label{tab:ICL_observer_results}
\begin{adjustbox}{angle=90}
\begin{tabular}{l c c c c c c c c}
\hline
Observer & $N_{\rm{tot}}$ &Mean ICL Frac&Mean (Obs - Sim)&Mean (Obs - Sim)&Mean (Obs - Sim)&$N>0$ &Mean (Obs - Sim)&  Mean Projection  \\
&&&30~kpc Aperture&50~kpc Aperture&100~kpc Aperture&&Kinematic& Scatter \\
\hline
\textit{Surface Brightness Threshold}\\
SB Martinez-Lombilla & 19 & 0.14 $\pm$ 0.04 & -0.23 $\pm$ 0.09 & -0.16 $\pm$ 0.08 & -0.07 $\pm$ 0.05 & 0 & -0.50 $\pm$ 0.16 & 0.06 $\pm$ 0.04 \\
Montes & 27 & 0.14 $\pm$ 0.03 & -0.21 $\pm$ 0.10 & -0.15 $\pm$ 0.08 & -0.07 $\pm$ 0.05 & 3 & -0.49 $\pm$ 0.13 & 0.07 $\pm$ 0.04 \\
\hline
Mean& & 0.14 $\pm$ 0.03 & -0.21 $\pm$ 0.10 & -0.16 $\pm$ 0.08 & -0.07 $\pm$ 0.05 && -0.50 $\pm$ 0.14 & 0.06 $\pm$ 0.04 \\
\hline
\textit{Composite Model}\\
Ahad & 50 & 0.12 $\pm$ 0.05 & -0.25 $\pm$ 0.13 & -0.19 $\pm$ 0.11 & -0.10 $\pm$ 0.07 & 2 & -0.53 $\pm$ 0.12 & 0.09 $\pm$ 0.07 \\
Ragusa & 34 & 0.14 $\pm$ 0.04 & -0.23 $\pm$ 0.12 & -0.17 $\pm$ 0.10 & -0.09 $\pm$ 0.07 & 2 & -0.50 $\pm$ 0.13 & 0.08 $\pm$ 0.04 \\
\hline
Mean & & 0.13 $\pm$ 0.05 & -0.24 $\pm$ 0.13 & -0.18 $\pm$ 0.10 & -0.09 $\pm$ 0.07 && -0.51 $\pm$ 0.13 & 0.09 $\pm$ 0.06 \\
\hline
\textit{Multi-Galaxy Fitting}\\
CICLE & 33 & 0.16 $\pm$ 0.05 & -0.19 $\pm$ 0.08 & -0.13 $\pm$ 0.06 & -0.05 $\pm$ 0.04 & 3 & -0.46 $\pm$ 0.14 & 0.10 $\pm$ 0.05 \\
\hline
\textit{Wavelet Decomposition}\\
DAWIS-SB & 61 & 0.13 $\pm$ 0.02 & -0.26 $\pm$ 0.17 & -0.19 $\pm$ 0.14 & -0.10 $\pm$ 0.10 & 5 & -0.53 $\pm$ 0.13 & 0.13 $\pm$ 0.11 \\
DAWIS-SS & 61 & 0.17 $\pm$ 0.08 & -0.21 $\pm$ 0.11 & -0.15 $\pm$ 0.09 & -0.05 $\pm$ 0.06 & 11 & -0.48 $\pm$ 0.15 & 0.22 $\pm$ 0.15 \\
DAWIS-W & 61 & 0.09 $\pm$ 0.02 & -0.29 $\pm$ 0.16 & -0.22 $\pm$ 0.13 & -0.13 $\pm$ 0.09 & 2 & -0.56 $\pm$ 0.13 & 0.17 $\pm$ 0.11 \\
\hline
Mean & & 0.13 $\pm$ 0.06 & -0.25 $\pm$ 0.15 & -0.19 $\pm$ 0.12 & -0.09 $\pm$ 0.09 && -0.52 $\pm$ 0.14 & 0.18 $\pm$ 0.13 \\
\hline
\hline
Overall Mean & & 0.13 $\pm$ 0.05 & -0.24 $\pm$ 0.13 & -0.18 $\pm$ 0.11 & -0.09 $\pm$ 0.08 && -0.51 $\pm$ 0.14 & 0.13 $\pm$ 0.11 \\
\hline
\end{tabular}
\end{adjustbox}
\end{table*}

\section{Discussion}
\label{sec:disc}
We have applied eight currently used observational measures (Surface Brightness Threshold, Non-Parametric Measures, Composite Models, Multi-Galaxy Fitting and Wavelet Decomposition) to mock images of 61 galaxy clusters from four cosmological hydrodynamical simulations. We then compared the BCG+ICL and ICL fractions obtained with the observational methods with those predicted in the simulations using five simulated measures (four aperture-based 0 - 1~Mpc, 30~kpc - 1~Mpc, 50~kpc - 1~Mpc, 100~kpc - 1~Mpc and one kinematic-based).  In this Section we explore some of the potential reasons for the differences we find between the observed and simulated BCG+ICL and ICL fractions and compare our results to earlier studies of ICL measurement fidelity.

\subsection{Considerations on the basic findings}
\label{sec:disc-cons}
We find mean observed BCG+ICL fractions of $0.51\pm0.12$ (Table~\ref{tab:BCGICL_observer_results}). Fig.~\ref{fig:ICLBCG_Frac_clustermass} shows that the BCG+ICL fractions, using any definition, range between 0.24 (Ahad) and 0.83 (DAWIS-SS), and within a single cluster the largest range is between 0.48 and 0.83 (Fig.~\ref{fig:Cluster_means}, left-hand panel). We find mean ICL fractions of $0.13\pm0.05$ (Table~\ref{tab:ICL_observer_results}). The range of ICL fractions, using any definition, ranges between 0.02 (DAWIS-W) and 0.34 (Ahad and DAWIS-SS; Fig.~\ref{fig:ICL_Frac_clustermass_aper}). Within a single cluster the largest range is between 0.11 and 0.34 (Fig.~\ref{fig:Cluster_means}, right-hand panel). 

\cite{Rudick2011} used a suite of N-body simulations of 6 galaxy clusters $0.8<M_{\odot} \times 10^{14}<6.5$ to measure the quantity of ICL using 5 methods from the literature (binding energy, kinematic separation, instantaneous density, density history and surface brightness threshold). They found that techniques that define the ICL solely based on the current position of the cluster luminosity, such as a surface brightness or local density threshold, tend to find less ICL than methods utilizing time or velocity information, including stellar particles' density history or binding energy. This was mainly because these measures did not include the ICL projected over the BCG. We also find that ICL fractions measured in apertures from simulations, or using any of the observed methods from the mock images (all methods based on the current position of the cluster luminosity), are significantly lower than those measured using the kinematic separation method. However, we do not find significant offsets in the Observed - Simulated aperture fractions as a result of the different observational methods for either the BCG+ICL or ICL fractions.

The range of ICL fractions found by \cite{Rudick2011} across all the clusters using any definition ranges between 9–36 per cent. Even within a single cluster, using different methods, the measured ICL fraction changes by up to a factor of two. 
The range of ICL fractions we find is slightly larger than found by \cite{Rudick2011}, and the range within a single cluster is larger. However, given that our sample includes ten times more clusters this is not unexpected. 

\cite{Kluge2021} explored four different methods to disentangle the BCG and ICL light in their observations of 170 galaxy clusters at $z<0.08$: a Surface Brightness Threshold ($\mu_{g'}<27$ mag/arcsec$^{2}$), excess light above a de Vaucouleurs profile \citep{devaucouleurs}, a composite model (2 S\'ersic profiles) and a luminosity cut ($M_g<-21.85$ mag). They find mean ICL fractions that vary from $0.10\pm0.12$ for the Surface Brightness Threshold, $0.13\pm0.09$ for excess light above a de Vaucouleurs profile, $0.18\pm0.17$ for the Composite Model and $0.20\pm0.12$ for the luminosity cut method. Our mean ICL fractions (Table~\ref{tab:ICL_observer_results}) are consistent with \cite{Kluge2021} within the uncertainties. However, we note that we do not observe an offset in measured ICL fraction depending on the observational measurement method employed; for the Surface Brightness Threshold, we obtain a mean fraction of $0.14\pm0.03$ and for the Composite Model of $0.13\pm0.05$. \cite{Kluge2021} also find a mean BCG+ICL fraction, over all methods, of $0.28\pm0.17$ which is lower than our mean of $0.51\pm0.12$. 

Compared to observational studies our BCG+ICL fractions are more consistent with the stacking analyses of \cite{Zibetti2005} ($0.33\pm0.16$) and \cite{Zhang2019} ($0.44\pm0.17$) and less consistent with \cite{Gonzalez2007} ($0.26\pm0.08$). Compared to simulations, our observationally-measured BCG+ICL fractions are similar to the fractions of \cite{Puchwein10} (0.45-0.59) and \cite{Proctor2023} ($\sim0.45$), lower than the fractions of \cite{Cui2014} (0.60-0.85) and consistent with the higher end of \cite{Contini2014} (0.23-0.61). This range suggests that there can be a dependence of ICL fraction on the simulations studied. This was also seen by \cite{Cui2014} who found that their ICL fractions changed by a factor of $1.5-2$ when they added AGN feedback to their simulations.

The BCG+ICL fraction primarily quantifies how well each method detects and excises the satellite galaxies in each image. The lower panel of Fig.~\ref{fig:ICLBCG_Frac_clustermass} shows that the observed BCG+ICL fractions are generally lower than the simulated measurements. The means of these values are given in Table~\ref{tab:BCGICL_observer_results} and the overall Mean is (Observed - Simulated) $=-0.05\pm0.09$. 

Fig~\ref{fig:ICL_Frac_ap_histo} shows that the observed ICL fractions are also generally lower than the simulated aperture measurements.  The difference is largest for the 30~kpc$ - 1$~Mpc aperture and decreases moving to the 100~kpc$ - 1$~Mpc aperture. The mean Observed-Simulated differences are given in Table~\ref{tab:ICL_observer_results}. The additional difference here compared to measuring the BCG+ICL fractions is a result of separating the ICL from the BCG. The observational measures are clearly more consistent with the 100~kpc - 1~Mpc aperture than with smaller inner radii. We explore this further in Section~\ref{sec:disc-rad}.

Fig~\ref{fig:ICL_Frac_clustermass_kin} shows that the simulated ICL fractions measured with the kinematic method are significantly larger than the observed fractions, with an overall mean (Observed-Simulated) $=-0.51\pm0.14$, as has been found previously by \cite{Rudick2011,Cui2014} and explored in more detail by \cite{remus17}.  Such significant differences suggest that observers and simulators are measuring very different quantities. We explore this further in Section~\ref{sec:disc-kin}.

A major difference between our work and earlier studies is the fact that here the different observational methods are each carried out by different people, each of whom applies different pre-processing steps before they make the measurements. We therefore test image processing and analysis methods as well as different ICL methods. The fact that the two measures that are the most similar in this analysis (SB Martinez-Lombilla and Montes) find an identical difference with the simulated 100~kpc$-1$~Mpc aperture ICL fraction (i.e. Observed-Simulated $= -0.07\pm0.05$) suggests that minor differences in approach (in this case circular vs elliptical apertures and binning vs not binning the images) from one observer to the next do not have a significant impact.

There are several reasons why there might be a lower BCG+ICL or ICL fraction in the observations than in the simulations. Projection effects may play a role: Image-based analyses mean that observers are working in a collapsed cylinder of radius 1~Mpc and length 4~Mpc (as a result of extracting the particles from a 4 Mpc cube, see Sec. 2.5), whereas the simulators work in a 1~Mpc radius sphere. There could also be an impact of the mass-to-light ratios applied to move from the simulation in mass units to the luminosity units of mock images. Light could also potentially be lost in the application of Gaussian noise to give the LSST-like surface brightness limit of $\mu_r=30.3$ mag/arcsec$^2$ in the creation of the mock images, as the simulations themselves are not limited in surface brightness, although we do not explore this further here. 

We considered whether the choice of cube size used to create the mock images affected the observers' measures. To explore this a few of the observers repeated their measurements on 4 mock images in a $2\times2$~Mpc cube, so the cylinder analysed became 1~Mpc radius and 2~Mpc length. The difference Fraction$_{\rm{4x4~Mpc}}$-Fraction$_{\rm{2x2~Mpc}}$ = $-0.008\pm0.100$~dex. This is not a significant offset but is another source of scatter.

\begin{figure}
	\includegraphics[width=\columnwidth]{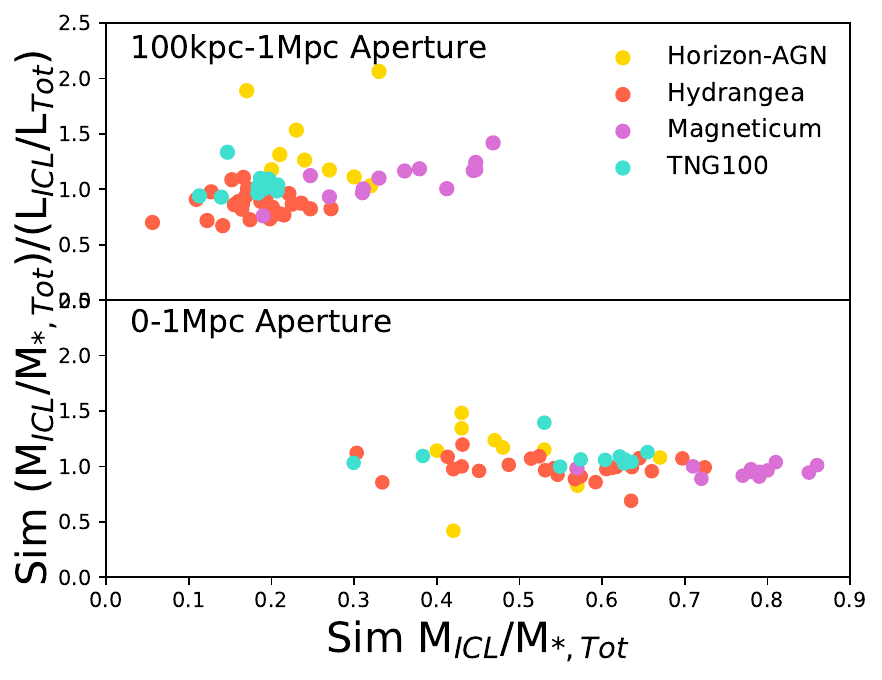} 
	\caption{Simulated BCG+ICL (upper panel) and ICL fraction (lower panel) of mass as a function of the fraction of luminosity compared to the simulated BCG+ICL or ICL fraction of mass. The different simulations are indicated in the legend.}
    \label{fig:Luminositymass}
\end{figure}

We also considered whether the fact that the different simulations use different star formation models which produce different metallicities and ages for their stellar particles causes offsets between the observed and simulated measurements because the observed luminosity fractions are measured from mock images (luminosity) whereas the simulated fractions are measured directly from the simulations (mass). While we are analysing fractions which will, to first order, divide out stellar population effects, and our mock image creation applies the same stellar population model to each of the simulations, different stellar particles having different stellar populations could imprint different mass-to-light ratios in the BCG compared to the ICL.  To explore this question, the simulation cubes were re-made applying a mass-to-light ratio to the stellar particles to create luminosity-based simulation cubes using the simple stellar population models used to create the mock images. The simulators re-measured their BCG+ICL and ICL fractions on the luminosity cubes.  Fig.~\ref{fig:Luminositymass} suggests that there are small, simulation-dependent offsets in the mass-light ratios. Table~\ref{tab:sim_lum_mass} gives the simulated ratios. (M$_{\rm{BCG+ICL}}$/M$_{*,\rm{Tot}}$)/(L$_{\rm{BCG+ICL}}$/L$_{\rm{Tot}}$) ranges from $\sim0.96\pm0.04$ for Magneticum to $1.10\pm0.29$ for Horizon-AGN. The simulated (M$_{\rm{ICL}}$/M$_{*,\rm{Tot}}$)/(L$_{\rm{ICL}}$/L$_{\rm{Tot}}$) is generally larger than for the BCG+ICL fraction and ranges from $\sim0.98\pm0.62$ for Hydrangea to $1.39\pm0.34$ for Horizon-AGN. The scatter around these ratios makes the values consistent with 1 and so this is not the cause of the systematic offsets observed between the observers and the simulators, rather it is an additional source of scatter in any comparison of simulated compared to observed BCG+ICL or ICL fractions.

Comparing between observations and simulations is often a significant element of ICL analyses \citep[e.g.][]{Montes2021} and the different simulations apply different star formation models which produce different metallicities and ages for their stellar particles. This analysis gives an idea of the scatter that these differences potentially introduce to that comparison. 

\subsection{Projection effects}
\label{sec:proj}
The differences we observe between the observations and simulations could be a result of the different projections that observers are measuring their fractions over. 

\begin{figure}
	\includegraphics[width=\columnwidth]{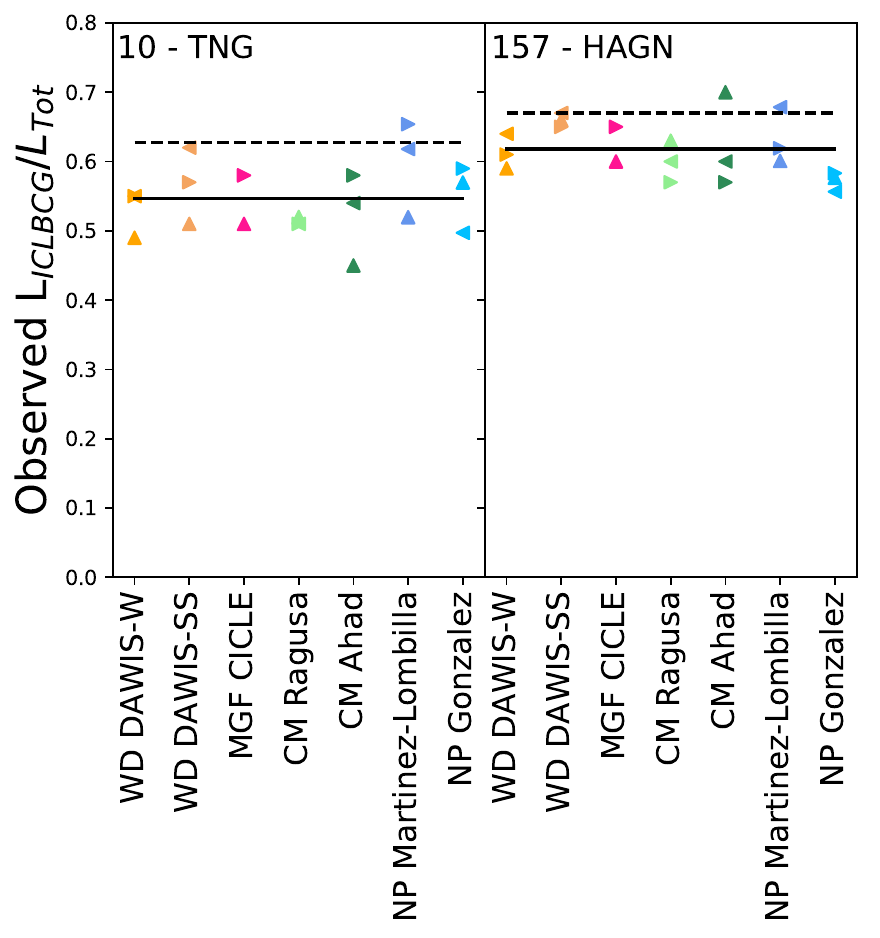} 
    \includegraphics[width=\columnwidth]{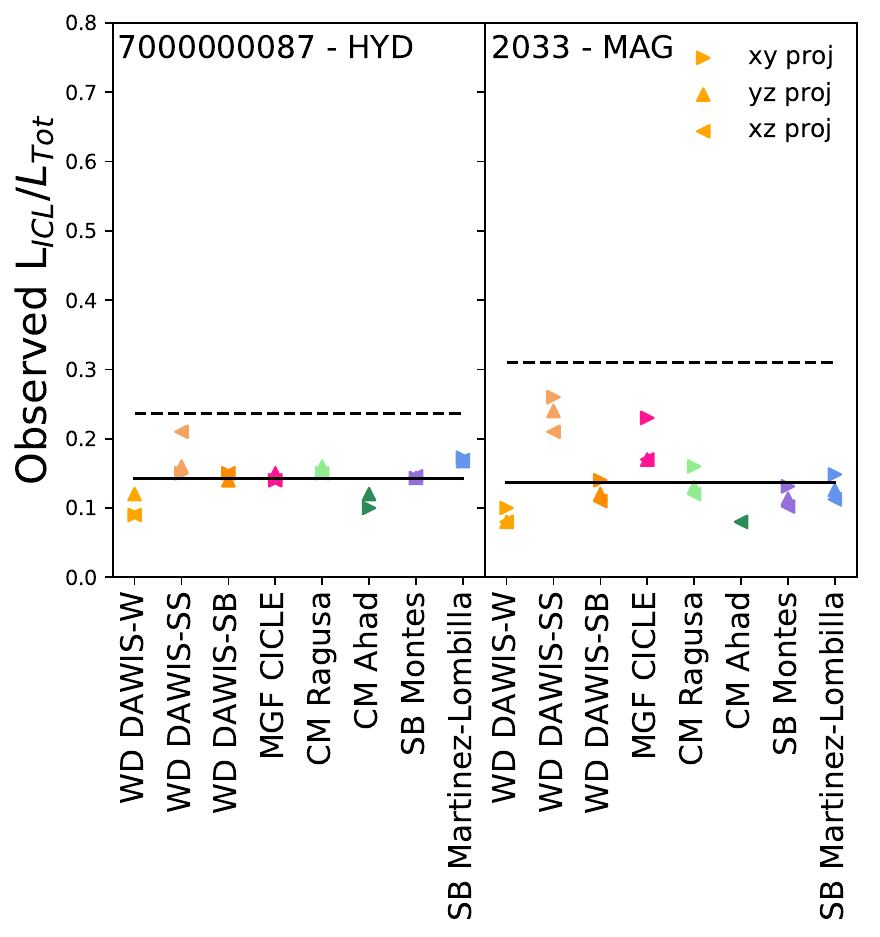}   
	\caption{Observed BCG+ICL (upper panel) and ICL (lower panel) fractions for each projection ($xy, xz, zy$ given by the legend) measured by each observer for clusters selected by the standard deviations around the mean observed fractions over all the measurements. The left-hand panels show clusters with small standard deviations (cluster 10 has mean BCG+ICL fraction $0.55\pm0.03$, Mean Projection Scatter = 0.07; cluster 7000000087 has mean ICL fraction $0.14\pm0.02$, Mean Projection Scatter = 0.22). The right hand panels show clusters with large standard deviations (cluster 157 has mean BCG+ICL fraction $0.56\pm0.08$, Mean Projection Scatter = 0.21; cluster 2033 has mean ICL fraction $0.15\pm0.06$, Mean Projection Scatter = 0.70). The cluster name and simulation are given in the top-left of each panel. The dashed lines give the simulated fractions ($0-1$~Mpc Aperture for the upper panels and 100~kpc - 1~Mpc Aperture for the lower panels) and the solid lines illustrate the mean observed fraction over all methods).}
    \label{fig:indiv_projections}
\end{figure}

Fig.~\ref{fig:indiv_projections} shows examples of the BCG+ICL (upper panel) and ICL (lower panel) fractions measured for 4 clusters, 2 with small (left-hand panels) and 2 with large (right-hand panels) standard deviations around the mean observed fractions over all the measurements. 

\begin{figure}
	\includegraphics[width=\columnwidth]{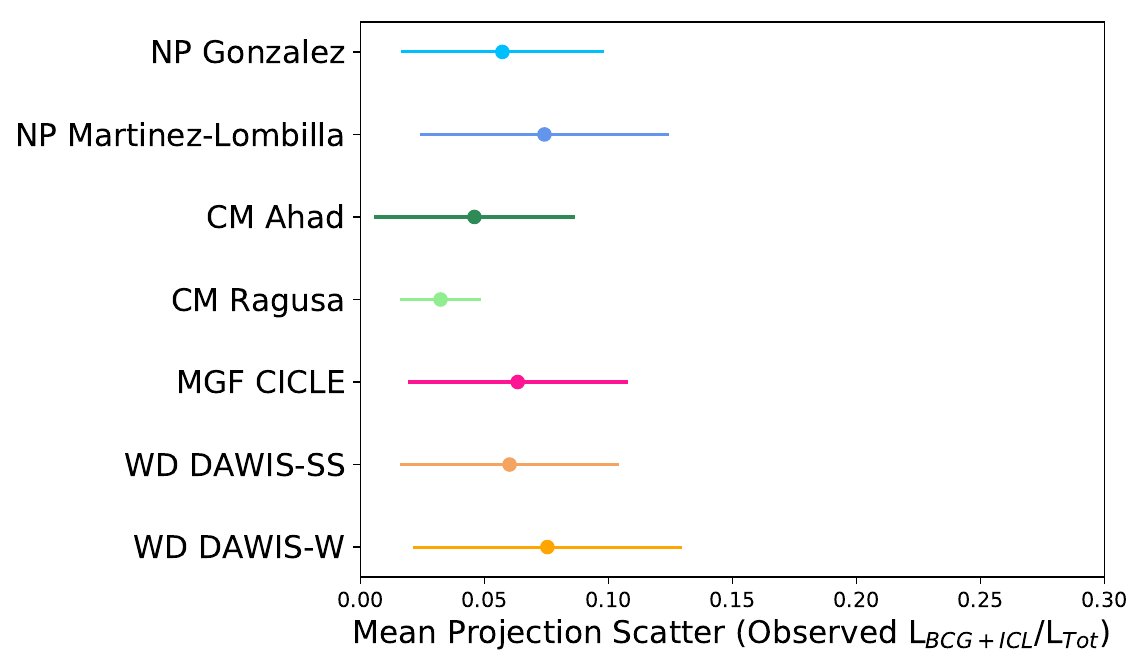}
    \includegraphics[width=\columnwidth]{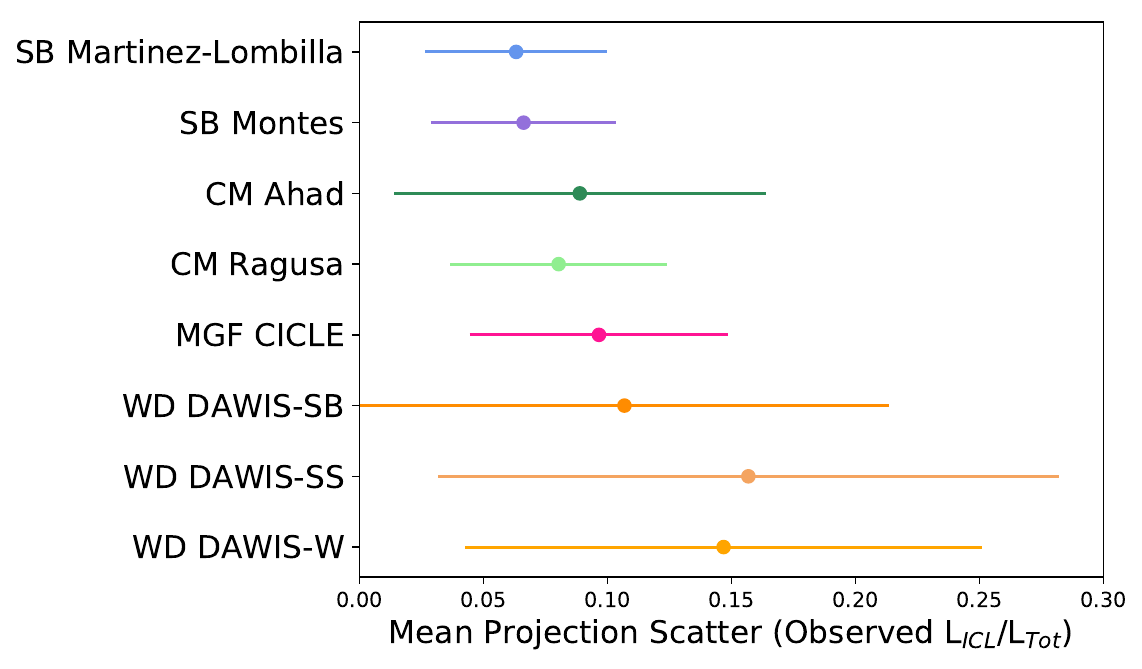}
    \caption{Mean Projection Scatter of BCG+ICL fraction (upper panel) and ICL fraction (lower panel) across the 3 different ($xy, xz, zy$) simulation projections for each of the observed measurements. The error bars indicate the standard deviation.}
    \label{fig:ICLBCG_Frac_rms}
\end{figure}

We calculate the `Projection Scatter' of the fractions over the projections ($xy, xz, zy$) measured for each cluster by each observer and scale this by the mean fraction calculated for that cluster by that observer, i.e., Projection Scatter = (Max$_{\rm{frac,cluster}}- $Min$_{\rm{frac,cluster}}$)/($2\,\times\,$Mean$_{\rm{frac,cluster}}$). We calculate half of the maximum-minimum range rather than the standard deviation as not all clusters have measurements for all 3 projections for each observer. We then calculate the mean and standard deviation around that `Projection Scatter' over all of the clusters measured by that observer. Fig.~\ref{fig:ICLBCG_Frac_rms} shows the Mean Projection Scatter for the BCG+ICL (upper panel) and ICL fractions (lower panel) for each observational method. There is significant scatter in the observed fractions between the different projections. The mean values for the BCG+ICL fractions are given in Table~\ref{tab:BCGICL_observer_results} and range from $0.03\pm0.02$ for Ragusa to $0.14\pm0.11$ for DAWIS-W, with an overall mean of $0.08\pm0.08$, i.e. an uncertainty from projection effects of 8 per cent. The lower panel of Fig.~\ref{fig:ICLBCG_Frac_rms} shows that the scatter as a result of projection effects is more significant for the observed ICL fractions. This is a result of the ICL fractions in the denominator of the Projection Scatter being smaller.  The mean values for the ICL fractions are given in Table~\ref{tab:ICL_observer_results}, with the differences ranging from $0.06\pm0.04$ for SB Martinez-Lombilla to $0.22\pm0.15$ for DAWIS-SS, and $0.13\pm0.11$ overall, an uncertainty from projection effects on ICL fractions of 13 per cent. The differences in Mean Projection Scatter between observers could be a result of their analysing different numbers of clusters (e.g. Fig.~\ref{fig:Cluster_means}). We tested this by measuring the Projection Scatter for a smaller sample of clusters that have BCG+ICL fractions measured by at least 6 out of 7 observers (26 clusters) and ICL fractions measured by at least 7 out of 8 observers (28 clusters). We find that the Mean Projection Scatter changed by at most 0.01, well within the standard deviations of the measurements. This suggests that the number of clusters analysed does not play a significant role in the differences in Mean Projection Scatter between observers. 

The observation-based analysis presented here often depends on the detection (and deblending) of the galaxies in the images. When observers were making their measurements, some found that some projections revealed galaxies (especially close to the BCG) that were unnoticed by \textsc{SExtractor} (or similar detection and/or deblending codes) in other projections. These undetected galaxies will play an important role in the final scatter displayed in Fig.~\ref{fig:ICLBCG_Frac_rms}, as well as in the number of clusters with Observed-Simulated $N>0$. Many of the techniques described here use \textsc{SExtractor} or similar to detect sources, so the ICL measurement problem is not only the separation of galaxy light from the ICL, but sometimes also galaxy detection itself. Another projection effect with a strong impact on some measurements (most notably DAWIS) is the apparent morphology of the cluster. As clusters appear to have different shapes under different projections, the simple criteria separations applied by the different DAWIS techniques here are not able to cohesively capture the range of cluster shapes, which results in large measurement scatters.

This analysis of the effects of projection shows that the randomness of projection effects produces larger uncertainties when trying to isolate the ICL than when isolating the BCG+ICL. The differences as a result of projection effects are consistent with the offsets between the observed and simulated measures for BCG+ICL fractions seen in Figs.~\ref{fig:ICLBCG_Frac_clustermass} but are not large enough to explain the offsets between the observed and simulated measures for the ICL fractions seen in Fig.~\ref{fig:ICL_Frac_clustermass_aper}. Projection effects clearly provide a potential source of scatter in the observed measurements, but do not explain the systematic offsets. 

\cite{Rudick2011} also examined the effect of 9 different viewing angles on their ICL fractions. They found that their ICL fractions varied by $\pm0.02$ on ICL fractions that ranged from 0.1 to 0.26. By our metric that is equivalent to projection effects of $8-20$ per cent, consistent with our findings of $\sim 13$ per cent. 

It is interesting that we find a similar offset of Observed-Simulated BCG+ICL and ICL fractions for each of the observational methods compared to the simulations. This suggests that the inherent difference of measuring ICL in different projections is significantly larger than the measurement scatter. In essence, we find that it is important to consider projection in any comparison between observations and simulations as it has a significant impact. 

\subsection{Where does the BCG end, and where does the ICL start?}
\label{sec:disc-rad}
One of the main questions when trying to define the ICL is where it starts to dominate. Without kinematic information available it is difficult to separate the BCG and the ICL to study the ICL and its evolution separately. However, many analyses have observed a change in the slope of the surface brightness profile of the BCG+ICL, suggesting that the ICL starts to dominate at that point.

\begin{figure}
	\includegraphics[width=\columnwidth]{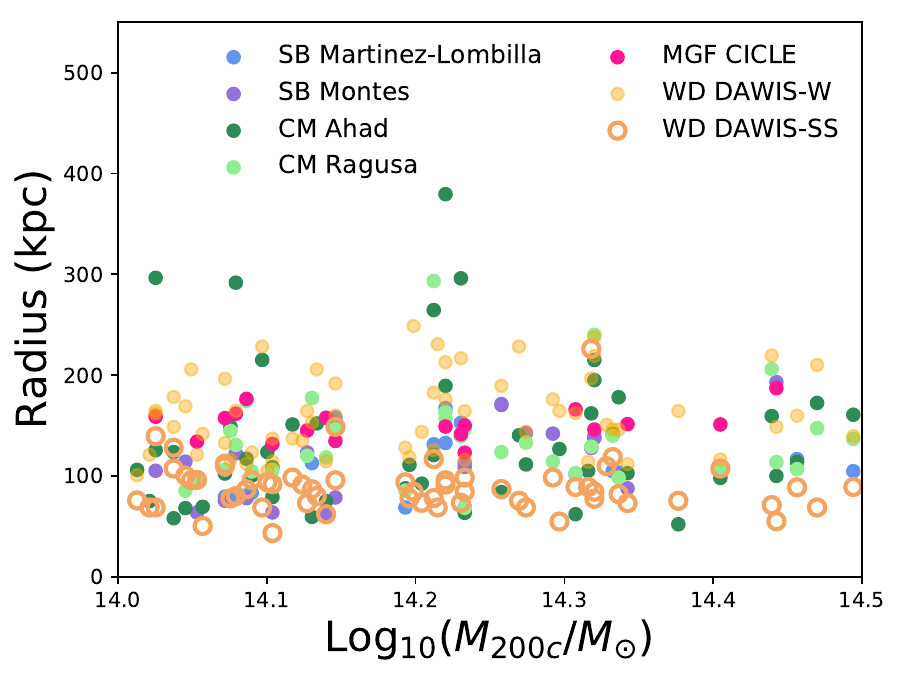}
    \includegraphics[width=\columnwidth]{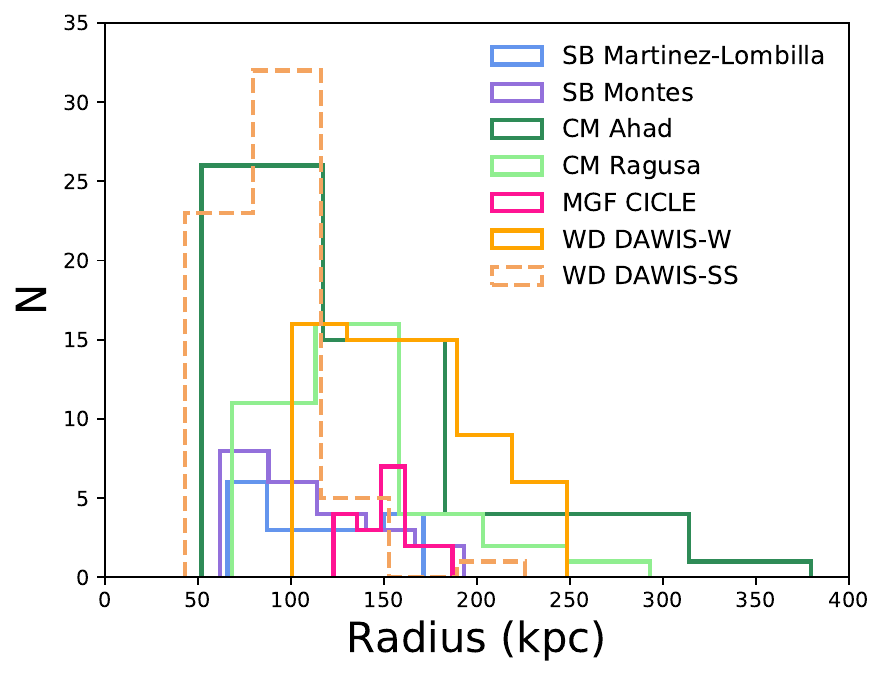}
    \caption{The radius (kpc) at which observers find the light to transition from BCG-dominated to ICL-dominated - The radius is the mean across the different ($xy, xz, zy$) projections in the simulations.  The upper panel shows the radius as a function of cluster mass and the lower panel shows a histogram of each observer's method. }
    \label{fig:Radius_clustermass}
\end{figure}

The observed ICL fractions are more consistent with the simulated 100~kpc - 1~Mpc aperture fractions than with the smaller inner apertures. Does this mean that the BCG-ICL transition is at 100~kpc for these clusters? Fig.~\ref{fig:Radius_clustermass} explores the radius at which the observers find the mock images to transition from BCG-dominated to ICL-dominated. The radius is the mean across the measured simulated projections. We note that some of the measured radii extend to much larger radii than the transition radii that observers have measured from observations to date (e.g. $\sim70-100$~kpc, \citealt{Montes2021}). We do not observe a relationship of radius with cluster mass over this mass range. Table~\ref{tab:Radii} gives the mean radii for each observer across the clusters they were able to measure. These mean radii vary from $89 \pm 31$~kpc for DAWIS-SS to $161 \pm 37$~kpc for DAWIS-W. The overall mean radius of $128 \pm 51$~kpc is consistent with the 100~kpc inner radius applied by the simulators. 

\begin{table}
\caption{Mean radius (kpc) at which observers find the mock images to transition from BCG-dominated to ICL-dominated. N is the number of clusters they were able to measure. Mean Radii for each observer. }
\label{tab:Radii}
\begin{tabular}{l c c}
\hline
Observer & N & Mean Radius (kpc) \\
\hline
\textit{Surface Brightness Threshold}\\
SB Martinez-Lombilla & 19 & 114 $\pm$ 33 \\
Montes & 23 & 110 $\pm$ 35 \\
\hline
\textit{Composite Model}\\
Ahad & 50 & 137 $\pm$ 70 \\
Ragusa & 34 & 136 $\pm$ 44 \\
\hline
\textit{Multi-Galaxy Fitting}\\
CICLE & 18 & 151 $\pm$ 15 \\
\hline
\textit{Wavelet Decomposition}\\
DAWIS-SS & 61 & 89 $\pm$ 31 \\
DAWIS-W & 61 & 161 $\pm$ 37 \\
\hline
\hline
Mean & & 128 $\pm$ 51 \\
\end{tabular}
\end{table}

\cite{chen22} investigated the BCG-ICL transition in a stacked image of 3000 clusters ($0.2<z<0.3$) in the SDSS $gri$ bands, and measured their BCG+ICL stellar surface mass profile down to 32~mag/arcsec$^{2}$ in the $r$-band. They decomposed the profile into three components, an inner de Vaucouleurs’ profile, an outer ICL that follows the dark matter distribution measured from weak lensing, and a transitional component between 70 and 200~kpc that represents the excess component in the diffuse light that cannot be described by the sum of a de Vaucouleurs’ profile and an ICL mass profile that follows the dark matter. They found that the ratio of the transitional component to the total diffuse mass peaks around 100~kpc. This could explain why we find a mean transition radius of $\sim128$~kpc.

\cite{contini2022} analysed their semi-analytic model to investigate the transition region between the BCG and the ICL. They defined this transition radius as the distance where the ICL accounts for 90 per cent of the total BCG+ICL mass. They found that the transition radius is independent of both BCG+ICL and halo mass and an average transition radius of 60~kpc, and as large as 100~kpc.

\cite{Proctor2023} used Gaussian Mixture Models to separate the ICL component in EAGLE simulations. They examined the transition radius where the ICL starts to dominate the stellar light, finding it to be $\sim100$ kpc for clusters of mass M$_{200c}\sim10^{14}M_{\odot}$, with a strong dependence on cluster mass.

While a 100 kpc aperture radius is not physically motivated, it appears that it is a reasonable approximation if we are comparing simulations with observations, as the ICL fractions in observations resemble those of the 100 kpc aperture in simulations for clusters with M$_{200c}\sim10^{14-14.5}M_{\odot}$ at $z\sim0$.

\subsection{Converting between image-based and kinematic ICL fractions}
\label{sec:disc-kin}

\begin{figure*}
  \includegraphics[width=0.9\textwidth]{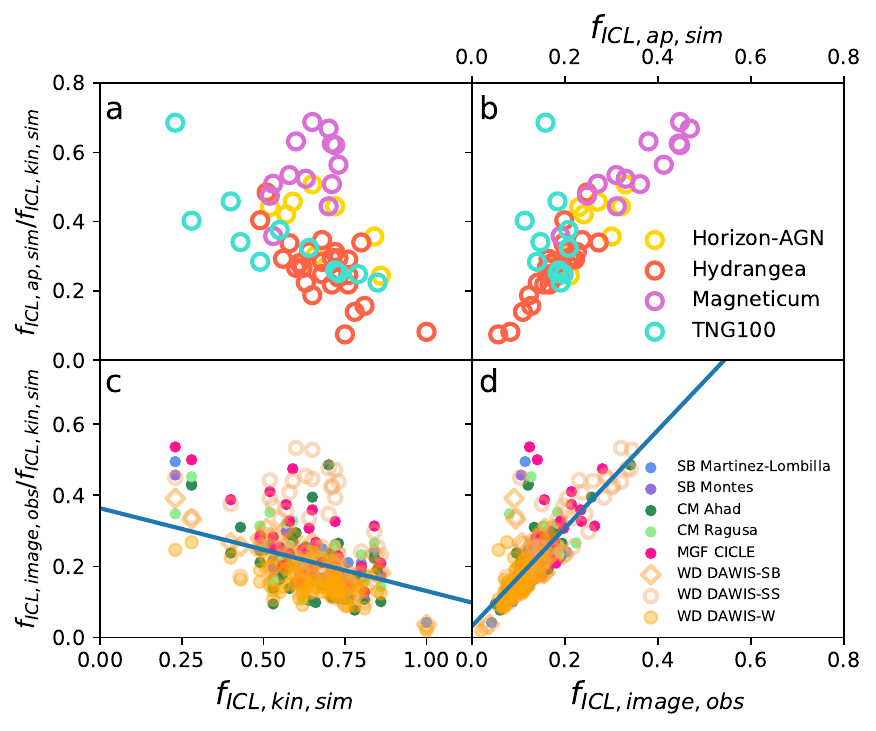}
    \caption{Relationship between ICL fractions measured by image-based and kinematic methods. Here we use the notation $f_{\rm{ICL,aperture,simulated}}$ and $f_{\rm{ICL,image,observed}}$ to indicate the image-based ICL fractions (simulated 100~kpc - 1~Mpc aperture and observed measures) and $f_{\rm{ICL,kinematic,simulated}}$ to indicate the kinematic ICL fractions (simulated measures). The top row (panels a and b) shows the simulated measurements and the bottom row (panels c and d) the observed measurements while the left panels (a and c) show the relationships as a function of the simulated kinematic ICL fraction and the right panels (b and d) show the relationships as a function of the image-based ICL fraction (simulated in b and observed in d). Estimating image-based or kinematic ICL fractions on the basis of the other is possible using the straight-line fits to the observed measures shown by the solid lines in the lower panels and given in Eqns.~\ref{eq:photom} (left panel) and \ref{eq:kin} (right panel).}
    \label{fig:SimObs_KinvsAp}
\end{figure*}

We explore how the image-based and kinematic ICL fractions are related in simulations and observations in Fig~\ref{fig:SimObs_KinvsAp}. Here we use the notation $f_{\rm{ICL,aperture,simulated}}$ and $f_{\rm{ICL,image,observed}}$ to indicate the image-based ICL fractions (simulated 100~kpc - 1~Mpc aperture and observed measures), and $f_{\rm{ICL,kinematic,simulated}}$ to indicate the kinematic ICL fractions (simulated measures). The kinematic ICL component is measured by decomposing the velocity component of the simulated galaxy clusters in 3D space using a double-Maxwellian distribution \citep{remus17}, very different from the image-based methods. The top row (panels a and b) of Fig~\ref{fig:SimObs_KinvsAp} shows the simulated measurements and the bottom row (panels c and d) the observed measurements, while the left panels (a and c) show the relationships as a function of the kinematic ICL fraction and the right panels (b and d) show the relationship as a function of the image-based ICL fractions. We find that the kinematic and image-based ICL fractions appear to be correlated in both the simulated and the observed measures. We calculate the Pearson correlation coefficients and find that the simulations relationships shown in Fig~\ref{fig:SimObs_KinvsAp} (panels a and b) have p-values of $0.038$ and $1.2\times10^{-16}$ respectively and the observed relationships (panels c and d) p-values of $1.9\times10^{-12}$ and $6.4\times10^{-70}$ respectively, showing that these are all significant correlations, with the exception of panel a which is only significant at a $\sim2\sigma$ level. We note that the Magneticum simulations can be seen to lie off the distribution in panel a. We re-calculate the correlation coefficients excluding these data. Without the Magneticum data the simulations relationships (panels a and b) have p-values of $0.001$ and $5.7\times10^{-6}$ respectively and the observed relationships (panels c and d) have p-values of $2.5\times10^{-16}$ and $8.9\times10^{-37}$ respectively, i.e. all of these relationships correlate significant at a $>3\sigma$ level.

Given that these relationships are significant, we are able to estimate the simulated kinematic ICL fraction from an observed, image-based, ICL fraction and vice versa. To achieve this we fit a straight line to all of the observed data, finding the relationships given in Eqns.~\ref{eq:photom} and \ref{eq:kin}, shown by the solid lines in panels c and d of Fig.~\ref{fig:SimObs_KinvsAp} respectively. The relationships described by Eqns.~\ref{eq:photom} and \ref{eq:kin} do not change beyond the uncertainties given when we re-fit them excluding the Magneticum data.

\begin{equation}
\label{eq:photom}
\frac{f_{\rm{ICL,image,obs}}}{f_{\rm{ICL,kin,sim}}}=(-0.23\pm0.03) f_{\rm{ICL,kin,sim}}+(0.36\pm0.02)
\end{equation}
\begin{equation}
\label{eq:kin}
\frac{f_{\rm{ICL,image,obs}}}{f_{\rm{ICL,kin,sim}}}=(1.38\pm0.06) f_{\rm{ICL,image,obs}}+(0.033\pm0.009)
\end{equation}
Re-arranging Eqn.~\ref{eq:photom} allows estimation of an observed, image-based, ICL fraction from the simulated kinematic ICL fraction and Eqn.~\ref{eq:kin} enables the reverse, estimating the simulated kinematic ICL fraction from an observed, image-based, ICL fraction. Further understanding how image-based and kinematic measurements of BCG+ICL and ICL fractions relate to one another is a very interesting problem that we will explore in greater detail in a later paper.

\subsection{Cluster relaxedness and BCG+ICL fraction}
\label{sec:disc-relax}

\begin{figure}
  \includegraphics[width=\columnwidth]{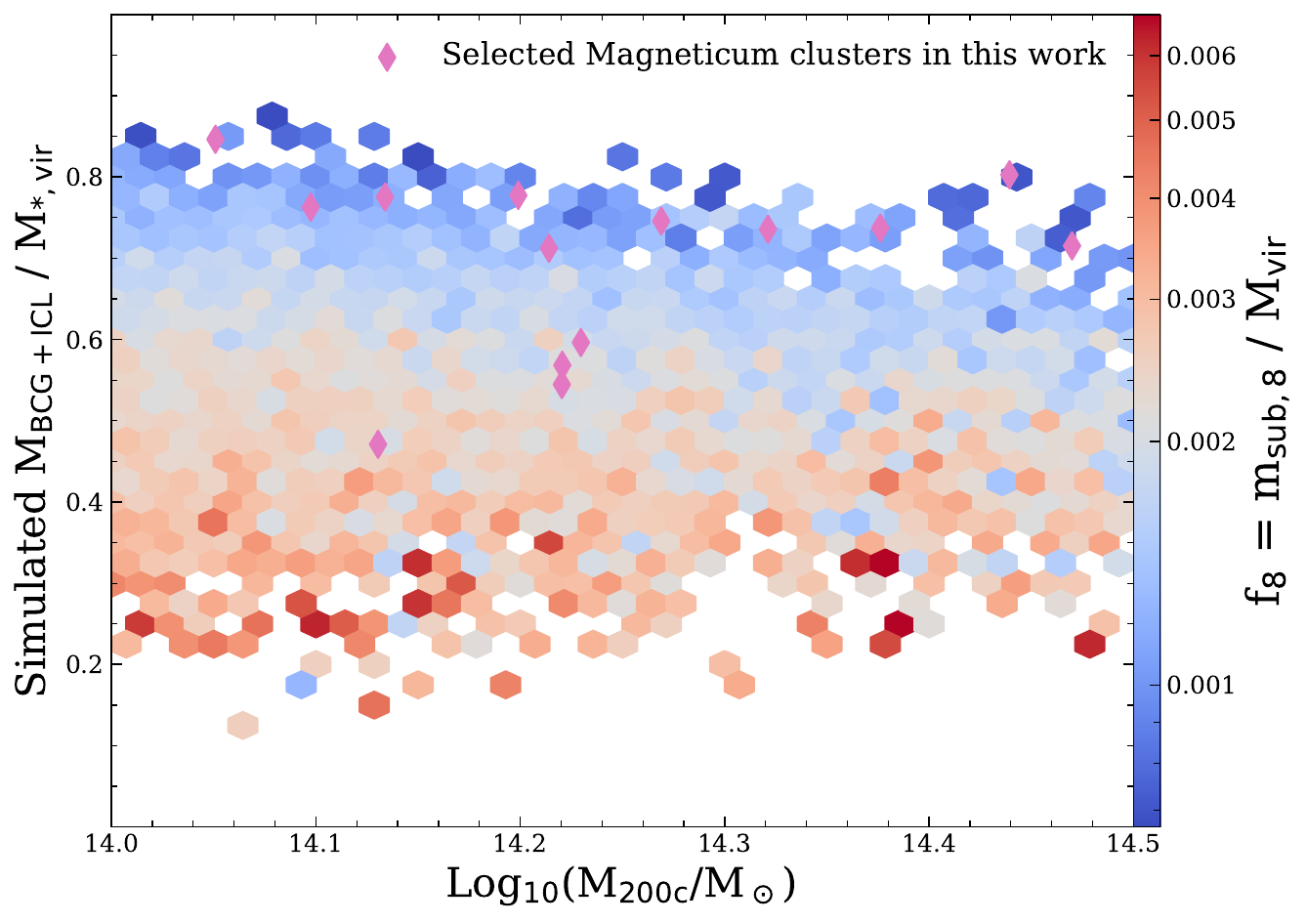}
    \caption{The fraction of mass in BCG+ICL as compared to the total stellar mass within the virial radius, r$_\mathrm{vir}$, for all clusters of Box2b of the Magneticum Pathfinder simulation, coloured by the mass fraction of the eighth subhalo, f$_\mathrm{8}$. Clusters with smaller f$_\mathrm{8}$ are more relaxed. Overlayed are the Magneticum clusters selected for this work (magenta diamonds), comprised primarily of very relaxed clusters.}
    \label{fig:ICLBCG_relaxed}
\end{figure}

As mentioned in Sec.~\ref{sec:select_clust}, the galaxy cluster sample from the large Magneticum simulation volume had to be selected in an automated fashion as there are too many galaxy clusters in the studied mass range to select them by hand. As the aim of this study is to test the separation of the ICL and BCG components, more relaxed clusters are generally preferable as they avoid scatter from the presence of significant substructure. Therefore, we applied a selection criterion that is an excellent tracer for the degree of relaxation of a galaxy cluster, namely the mass fraction of the 8th most massive substructure, $f_8$, with smaller values of $f_8$ denoting more relaxed clusters, because in a relaxed cluster the substructures are less prominent \citep[see][for more details]{kimmig22}.

Fig.~\ref{fig:ICLBCG_relaxed} shows the simulated measurement of the BCG+ICL fraction relative to the total stellar mass of the galaxy cluster for the full sample of more than 1000 galaxy clusters from the Magneticum Box2b, as a function of their total halo mass $M_\mathrm{200c}$, stacked and colour-coded according to the average $f_8$ within a bin. The BCG+ICL fractions of the full Magneticum sample are consistent with our finding that there is no trend of the BCG+ICL fraction with cluster mass. However, there is a clear trend for the BCG+ICL fraction to be higher for more relaxed galaxy clusters (up to 80 per cent, bluer symbols), and smaller for clusters that are currently assembling (down to 20 per cent, redder symbols). This might be expected, as more relaxed galaxy clusters have had more time to disrupt the accreted galaxies and add their stellar content to the BCG and/or the ICL. Note, however, that there is scatter in this relation so clusters that are currently assembling with large BCG+ICL fractions and vice versa also exist. 

The galaxy clusters selected for this study are shown as magenta diamonds in Fig.~\ref{fig:ICLBCG_relaxed}. Most of the selected clusters are at the uppermost range of BCG+ICL fraction compared to the full galaxy cluster sample, as expected due to their selection as relaxed systems. This explains the origin of the large BCG+ICL fractions found for the Magneticum clusters in comparison to the other simulations included in this study, as seen in Fig.~\ref{fig:Sims_bcgmass}. The middle panel of Fig.~\ref{fig:Sims_bcgmass} also shows that the lowest BCG+ICL fraction found for the Magneticum sample marks the most relaxed cluster from Box4, which is still somewhat unrelaxed as the volume of the simulation is too small to harbour large relaxed clusters at $z=0$. This system can be seen to be consistent with the BCG+ICL fractions for the other simulations which have similar box sizes (Table~\ref{ap:table_sims}). As the BCG+ICL fraction for this cluster agrees well with similar clusters (in mass and relaxation) from the less resolved Magneticum simulation volume (Box2b), this provides confidence that any differences are not a result of the different simulation resolutions.

To test whether the increased scatter added to the analysis by including the relaxed Magneticum clusters affects our ability to distinguish between different measurement methods, we repeated our analysis without those 14 systems. We found that the mean Observed-Simulated BCG+ICL fractions for each method reduced by at most 0.01, with no change in the standard deviation around those. The mean overall Observed-Simulated BCG+ICL fractions was $-0.05\pm0.06$ (a change from $-0.06\pm0.06$). The mean Observed-Simulated (100~kpc - 1~Mpc aperture) ICL fractions for each method changed a little more with reductions of 0.01-0.02 in the fractions and 0.01-0.04 in the standard deviations. This resulted in a mean overall Observed-Simulated ICL fraction of $-0.06\pm0.04$ (a change from $-0.09\pm0.07$). While this did reduce the scatter for each method, all the measurement methods changed similarly, meaning that even when excluding the Magneticum clusters we find no evidence of an offset in measured BCG+ICL or ICL fraction as a result of the observational method employed.

Given the impact of cluster relaxation on BCG+ICL and ICL fractions seen in Figs.~\ref{fig:Sims_bcgmass} and~\ref{fig:ICLBCG_relaxed}, we will explore this question in more detail in a later paper. 

\subsection{On the scalability of the observational methods}

How easy is it to apply each of these observational methods to large amounts of data, like that shortly to be available for the LSST \citep[e.g.][]{Montes2019,Brough2020} and Euclid surveys \citep[e.g.][]{Euclid2022,EuclidCol2022}? The Surface Brightness threshold is the easiest and simplest method to apply and in this particular case, where there are no foreground stars or background galaxies, does not require masking. It does not assume a particular morphology for the ICL but the threshold itself can vary as a result of different photometric bands and redshifts, leading to different ICL fractions, making this a challenging method to compare between studies.

The Composite Model methods require masking of the data. This process often still requires manual intervention which leads to the lower number of clusters analysed using those methods presented here.  However, our analysis shows that these methods are among the most consistent with the simulations (considering the 100~kpc - 1~Mpc aperture ICL fraction measure) and show a similar level of uncertainty from projection effects ($6$ (NP) and $4$ per cent (CM) for BCG+ICL Fraction and $6$ (SB) and $9$ per cent (CM) for ICL fraction). The Ahad composite model measure presented here has been designed to analyse large numbers of systems and does mask automatically \citep{Ahad2023}. We find it to give good fidelity compared to the simulation measures and a similar level of uncertainty with respect to projection effects. It does have a slightly larger standard deviation around the Mean Projection Scatter for ICL fraction, but that may be expected as we move to more automated measures.

The CICLE multi-galaxy fitting and DAWIS wavelet decomposition algorithms are easier to run on larger samples than any but the Ahad algorithm. We find the CICLE multi-galaxy fitting algorithm provides fractions that are closest to the simulations (Table~\ref{tab:BCGICL_observer_results} and~\ref{tab:ICL_observer_results}), for both the BCG+ICL and the ICL fractions. Additionally, the impact of projection effects on CICLE is comparable to that of the Non-parametric and Composite Model techniques. The DAWIS wavelet decomposition fractions have a similar low offset but more scatter compared to the simulation measures and we find a much higher level of uncertainty from projection effects with the current iteration of the algorithm used in this analysis.

It is only very recently that we have had the prospect of enough ICL observations that ICL-measurement algorithms have needed to be scalable to large samples. This exercise shows that effort is still needed to make existing algorithms scalable and to tune new methods to have the fidelity of earlier methods. We will make a subset of these data publicly available to provide a standard dataset to check the fidelity of new BCG+ICL or ICL measurement algorithms as they are developed.

\section{Conclusions}
\label{sec:concl}
We have applied eight currently used observational measures (Surface Brightness Threshold, Non-Parametric Measures, Composite Models, Multi-Galaxy Fitting and Wavelet Decomposition) to mock images of 61 galaxy clusters from four of the most widely used cosmological hydrodynamical simulations (Horizon-AGN, \citealt{Dubois.2014};  Hydrangea, \citealt{Bahe_et_al_2017}; Illustris-TNG, \citealt{Nelson2019}; and Magneticum, \citealt{dolag17}). We then compared the results obtained with the observational methods with the amount of ICL predicted in the simulations using five simulated measures (four aperture-based 0 - 1~Mpc, 30~kpc - 1~Mpc, 50~kpc - 1~Mpc, 100~kpc - 1~Mpc and one kinematic-based). From this analysis we conclude the following.
\begin{itemize}
\item{On average the different simulations give more or less consistent BCG+ICL (overall mean $0.58\pm0.14$) and ICL (overall mean $0.38\pm0.16$) fractions, with the exception of Magneticum whose higher fractions are shown to be a result of selecting very relaxed clusters from this simulation.}
\item{The different observational techniques give surprisingly consistent BCG+ICL (overall mean $0.51\pm0.12$) and ICL (overall mean $0.13\pm0.05$) fractions. The different observational techniques all tend to be biased to underestimating the BCG+ICL (mean difference $-0.05\pm0.09$) and ICL (mean difference $-0.09\pm0.08$, 100~kpc - 1~Mpc aperture) fractions when compared with measurements from the simulations. We find that the simulated 100~kpc - 1~Mpc aperture fraction is the most consistent with the observed ICL fractions, among all radii considered in this work.}
\item{The values of the ICL fractions measured by kinematic separation are significantly larger than the observed fractions with an overall mean (Observed-Simulated) $=-0.51\pm0.14$.  Such significant differences suggest that observers and simulators are measuring two very different quantities. We find that these measurements are related and offer fitted relations (Eqs.~\ref{eq:photom} and \ref{eq:kin}) to enable observers to estimate kinematic ICL fractions from image-based measurements and vice versa.}
\item{Exploring the reasons for the offsets in the Observed-Simulation fractions we do not find a single source for the offset but we do find several potential sources of increased measurement uncertainty: (1) The choice of cube size containing the cluster has a minor impact on measurement uncertainty. (2) The different simulations use different star formation models which give their stellar particles different ages and metallicities, this adds an additional source of scatter in any comparison of simulated compared to observed BCG+ICL or ICL fractions. (3) Projection effects are substantial in these measurements and cause uncertainties of 3-14 per cent (overall mean $0.08\pm0.08$) for BCG+ICL fractions and 6-22 per cent for ICL fractions (overall mean $0.13\pm0.11$).}
\item{We find a mean observed BCG-ICL transition radius of $128\pm51$~kpc.  Simulated ICL fractions measured by excising an inner radius of 100~kpc appear to be a reasonable approximation for the image-based ICL fraction in clusters with $10^{14-14.5}\rm{M_{\odot}}$ at $z\sim0$ when compared with measurements from observations. Therefore, to separate the ICL and obtain a robust quantity for analysis, 100~kpc is the minimum inner radius that one would wish to use for such a separation.}
\item{Comparing the different methods and algorithms, we note that measuring the combined BGC+ICL fraction has the least bias and scatter and is least affected by projection effects of all of the measures. The Surface Brightness Threshold, Non-Parametric Measure and Composite Model methods are among the methods most consistent with the simulations and show the lowest impact from projection effects. However, they have some known issues and only one of the algorithms tested here is set up for analysing larger samples ($N>50$) by masking automatically \citep{Ahad2023}. The CICLE multi-galaxy fitting \citep{Jimenez-Teja2016} and DAWIS wavelet decomposition \citep{Ellien2019} algorithms are better set up for analysing larger samples. CICLE is most consistent with the simulations, and has projection effects that are comparable to the Non-Parametric Measure and Composite Model methods. However, the measured fractions from DAWIS have more scatter compared to the simulation measures and a higher level of uncertainty from projection effects in the algorithm's current form.}
\item{We recommend that new algorithms be explored based on these methods to respond to the influx of data from the next generation of imaging surveys like LSST and Euclid. To assist with this we make a subset of these data publicly available to provide a standard dataset to test new BCG+ICL or ICL measurement algorithms as they are developed.}
\item{Due to the uncertainties we find induced by projection effects and the different ways of measuring BCG+ICL and ICL fractions, we suggest that authors not constrain models with analyses of single clusters at low redshift. We also recommend that relationships examined using homogeneous samples and methods will be significantly more robust than more heterogeneous comparisons.}
\end{itemize}

The caveat to all of these conclusions is the effects of cluster mergers on these measurements. We also note that no observational bias has been taken into account in our analysis. We will explore these and related questions in future papers.

\section*{Acknowledgements}

We thank the anonymous referee for their very constructive comments that have improved the final paper. SB acknowledges funding support from the Australian Research Council through a Discovery Project DP190101943. YB acknowledges funding from the Dutch Research Organisation (NWO) through Veni grant number 639.041.751 and financial support from the Swiss National Science Foundation (SNSF) under project 200021\_213076. Y.J-T. acknowledges financial support from the European Union’s Horizon 2020 research and innovation programme under the Marie Skłodowska-Curie grant agreement No 898633, the MSCA IF Extensions Program of the Spanish National Research Council (CSIC), the State Agency for Research of the Spanish MCIU through the Center of Excellence Severo Ochoa award to the Instituto de Astrofísica de Andalucía (SEV-2017-0709), and grant CEX2021-001131-S funded by MCIN/AEI/ 10.13039/501100011033. MM acknowledges the Project PCI2021-122072-2B, financed by MICIN/AEI/10.13039/501100011033, and the European Union “NextGenerationEU”/RTRP and IAC project P/302302. . This work used the DiRAC@Durham facility managed by the Institute for Computational Cosmology on behalf of the STFC DiRAC HPC Facility (www.dirac.ac.uk). The equipment was funded by BEIS capital funding via STFC capital grants ST/K00042X/1, ST/P002293/1 and ST/R002371/1, Durham University and STFC operations grant ST/S003908/1. DiRAC is part of the National e-Infrastructure. LK acknowledges support by the COMPLEX project from the European Research Council (ERC) under the European Union’s Horizon 2020 research and innovation program grant agreement ERC-2019-AdG 882679.

%%%%%%%%%%%%%%%%%%%%%%%%%%%%%%%%%%%%%%%%%%%%%%%%%%
\section*{Data Availability}

The Horizon-AGN data used in this work can be obtained upon request from https://www.horizon-simulation.org/data.html.

The Hydrangea simulation data are publicly available at https://ftp.strw.leidenuniv.nl/bahe/Hydrangea/.

The IllustrisTNG data used in this work are publicly available at http://www.tng-project.org.

Magneticum data are partially available at https://c2papcosmosim.uc.lrz.de/ \citep{ragagnin17}, with larger data sets on request.

%%%%%%%%%%%%%%%%%%%% REFERENCES %%%%%%%%%%%%%%%%%%

% The best way to enter references is to use BibTeX:

\bibliographystyle{mnras}
\bibliography{icl} % if your bibtex file is called icl.bib

%%%%%%%%%%%%%%%%% APPENDICES %%%%%%%%%%%%%%%%%%%%%

\appendix

\section{Simulation Table}
\begin{table*}
\scriptsize
\begin{center}
    \caption{Cosmological (magneto-)hydrodynamical simulations of massive clusters of galaxies adopted in this work. Here we include only cosmological models, i.e. simulations that start from cosmologically-motivated initial conditions on large spatial scales, which are run to $z\sim0$. These simulations differ in that they adopt not only different codes (Smooth-Particle-Hydrodynamics, Adaptive-Mesh-Refinement, meshless or moving mesh) but also different underlying galaxy formation models. All simulations include feedback from Super-Massive Black Holes, but with varying choices and implementations. IllustrisTNG includes MHD. Magneticum includes thermal conduction.}
    \label{tab:sims}
    
    \begin{tabular}{l c c c c }
    \hline
    Simulation project & Hydrangea & Horizon-AGN & Magneticum & IllustrisTNG \\
    \hline
    
    Run(s) & Hydrangea Zooms & AGN & Box4, Box2b & TNG100 \\
    Code & GADGET-3 & RAMSES & GADGET-3 & AREPO \\
    Lowest available redshift & $z=0$ & $z=0$& $z=0.2$&$z=0$ \\
    Box Size [com Mpc] & 3200$^a$ & 142 & 68, 909  & 111 \\
    Star-particle Mass Resolution [$10^6 \MSUN$] & 1.8 & 2.0 & 2.6, 50 & 1.4 \\
    & & & & \\
    \# clusters with $M_\mathrm{200c}\ge 10^{14}\,\MSUN$ & 24 & 14& 3, 4268 & 14\\
    \# clusters analyzed in this paper$^b$ & 27 & 14& 1, 13 & 11\\
    & & & & \\
    $\Lambda$CDM Cosmology & Planck2014 & WMAP7 & WMAP7& Planck2015 \\
     & \citet{Planck2014} & \citet{komatsu11} &\citet{komatsu11}& \citet{Planck2016} \\
    %Gas Mass Resolution [$10^6 \MSUN$] & $1.8$ & grid (stellar $2\times10^{6}~{\rm M_{\odot}}$) & ? & $1.4 \times 10^{6}$ \\
    & & & & \\
    Star formation & density threshold & density-threshold & density-threshold & density-threshold \\
    & & & & \\
    Stellar feedback: method & direct ISM heating & direct (momentum and energy) & direct energy, temporary & temporary hydro decoupling \\
    & & &  decoupled momentum & \\
    Stellar feedback: timing & stochastic, $\Delta T=10^{7.5}K$ & continuous (winds + SNII + SNIa)$^{\dag}$ & (continuous thermal, probabilistic  & continuous probabilistic, $\propto$ SFR \\
    & & & winds) $\propto$ SNII, & \\
    & & & continuous thermal $\propto$ SNIa & \\
    Stellar feedback: feedback   &  thermal & kinetic + thermal & kinetic + thermal & kinetic + thermal (warm)\\
    Stellar feedback: orientation & random & isotropic & isotropic & isotropic \\
    & & & & \\
    SMBH: seed mass [$10^6 \MSUN$] & & 0.1 & 0.12, 0.45 & 1.2\\
    SMBH: accretion & & Eddington/Bondi-Hoyle-Lyttleton & Eddington/Bondi-Hoyle-Lyttleton & Bondi–Hoyle\\
    SMBH feedback: mode(s) & thermal & thermal (high), kinetic (low) & dual: radio/quasar mode$^*$& dual:high-state/ low-state \\
    SMBH feedback: timing & stochastic, $\Delta T=10^{9}K$ & continuous & contineous & continuous/pulsated\\
    SMBH feedback: energy & thermal  & thermal/kinetic & thermal & thermal/kinetic\\
    SMBH feedback: orientation & random & isotropic (high) / bipolar (low) & isotropic & isotropic\\
    & & & & \\
    Simulation/Method References & \cite{Schaye.2015} & \cite{Dubois.2014} &  \cite{hirschmann14} & $\clubsuit$ \\
    & \cite{Bahe_et_al_2017} & & \cite{teklu15} & \\
    \hline
    \end{tabular}   
\end{center}
 
$^a$ Here the box size denotes the size of the parent box: Hydrangea comprises a number of so-called zoom-in simulations, with haloes identified and resimulated out of a large parent box. \\
$^b$ For this paper, we focus on clusters in a narrow mass range, namely:  $\log_{10}\,(M_\mathrm{200c}\,/\MSUN{}) = [14.0, 14.5]$. Additionally, in the case of the Magneticum run Box2b, we apply additional selection criteria based on relaxedness (see text for details).\\
$\dag$ SNII: \citep{Girardi2000}, winds: \citep{Leitherer92}, SNIa: \citep{Matteucci86}  \\
$^*$ \cite{fabjan10}\\
$\clubsuit$ \cite{Pillepich2018b, Nelson2018, Springel2018, Marinacci2018,Naiman2018, Nelson2019}\\

\label{ap:table_sims}
\end{table*}

Table \ref{ap:table_sims} gives a summary of the main parameters of the different cosmological simulations used in this work. 

\section{Fractions per Cluster}
\label{ap:clusters}
\begin{figure*}
	\includegraphics[width=\columnwidth]{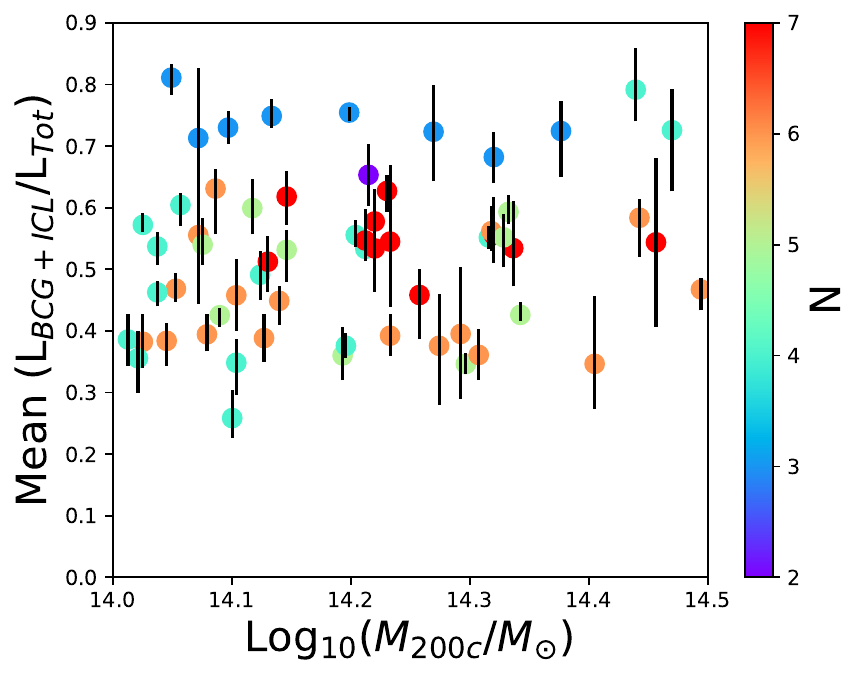}
    \includegraphics[width=\columnwidth]{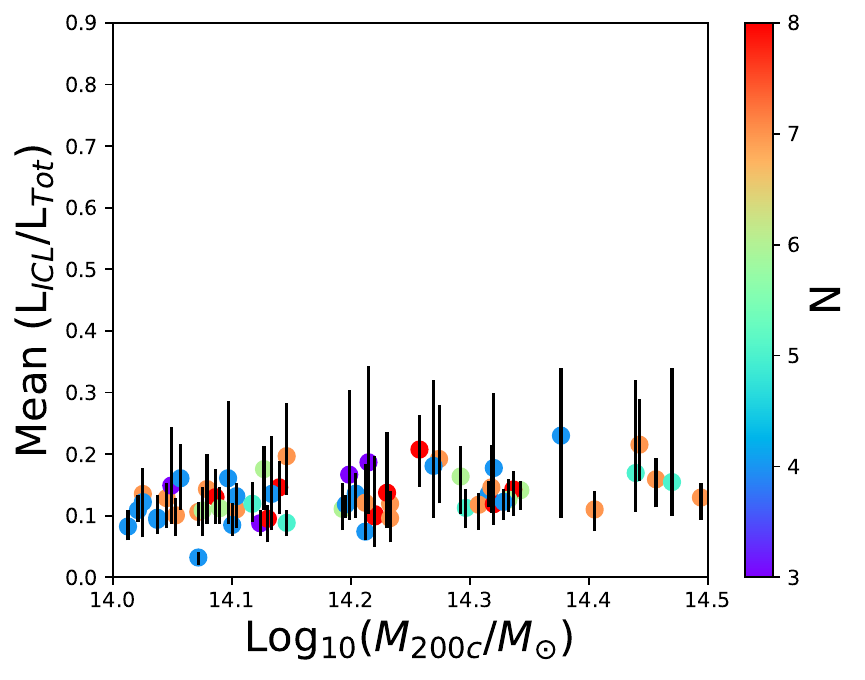}
    \caption{The mean BCG+ICL (left panel) and ICL (right panel) fractions averaged over all measures as a function of cluster mass. The colours indicate the number of measurements made for each cluster. The error bars indicate the minimum and maximum fraction measured for each cluster. }
    \label{fig:Cluster_means}
\end{figure*}

Fig. \ref{fig:Cluster_means}
 shows the average of all the observed BCG+ICL (left) and ICL (right) fractions per cluster, as a function of cluster mass. The measurements are colour-coded by the number of individual observer measurements per cluster. This shows that the average fractions do not depend on the number of measurements included in the average.
 
%%%%%%%%%%%%%%%%%%%%%%%%%%%%%%%%%%%%%%%%%%%%%%%%%%

% Don't change these lines
\bsp	% typesetting comment
\label{lastpage}
\end{document}